\documentclass[]{aa}
\usepackage[varg]{txfonts}
\usepackage{graphicx}
\usepackage{booktabs}
\usepackage{amsmath}
\usepackage{colortbl}
\usepackage{subfig}
\usepackage{rotating}
\usepackage{amssymb}
\usepackage{latexsym}
\usepackage{ifthen}
\usepackage{enumitem}
\usepackage{mathtools}
\usepackage{mathrsfs}   
\usepackage{xcolor,soul,lipsum}

\usepackage[hidelinks]{hyperref}

\hypersetup{%
  colorlinks=true,
  linkcolor=blue,
  allcolors=blue
}

\begin{document}

\title{Multi-phase outflows in Mkn 848 observed with SDSS-MaNGA Integral Field Spectroscopy} 
\titlerunning{Multi-phase outflows in Mkn 848} 
\author{M. Perna
                \inst{\ref{i1}}\thanks{E-mail: michele.perna@inaf.it}
                \and 
        G. Cresci
                \inst{\ref{i1}}        
                \and
        M. Brusa
                \inst{\ref{i2},\ref{i3}}
                \and
        G. Lanzuisi
                \inst{\ref{i2},\ref{i3}} 
                \and
       A. Concas
                \inst{\ref{i4},\ref{i5}}\thanks{Current affiliation:  Kavli Institute for Cosmology, University of Cambridge, Madingley Road, Cambridge CB3 0HA, UK} 
                \and
       V. Mainieri
                \inst{\ref{i6}} 
                \and
       F. Mannucci
                \inst{\ref{i1}} 
                \and
       A. Marconi
                \inst{\ref{i1},\ref{i7}}        
}

\institute{INAF - Osservatorio Astrofisico di Arcetri, Largo Enrico Fermi 5, I-50125 Firenze, Italy\label{i1}
        \and
        Dipartimento di Fisica e Astronomia, Universit\`a di Bologna, via Gobetti 93/2, 40129 Bologna, Italy\label{i2}
        \and
        INAF - Osservatorio di Astrofisica e Scienza dello Spazio di Bologna, via Gobetti 93/3, 40129 Bologna, Italy\label{i3}
        \and
        Excellence Cluster Universe, Boltzmannstr. 2 D-85748 Garching, Germany\label{i4}
        \and
        Technische Universit\"at Munchen, Physik-Department, James-Frank-Str. 1, D-85748, Garching bei M\"unchen, Germany\label{i5}
        \and
        European Southern Observatory, Karl-Schwarzschild-Str. 2, 85748, Garching bei M\"unchen, Germany\label{i6}
        \and
        Dipartimento di Fisica e Astronomia, Universit\`a degli Studi di Firenze, Via G. Sansone 1, 50019 Sesto Fiorentino, Firenze, Italy\label{i7}
}

\date{Received 5 September 2018 / Accepted 18 December 2018}

\abstract {} 
{
The characterisation of galaxy-scale outflows in terms of their multi-phase and multi-scale nature, amount, and effects of flowing material is crucial to place constraints on models of galaxy formation and evolution. This study can proceed only with the detailed investigation of individual targets.
} 
{
We present a spatially resolved spectroscopic optical data analysis of Mkn 848, a complex system consisting of two merging galaxies at z $\sim 0.04$ that are separated by a projected distance of 7.5 kpc. Motivated by the presence of a multi-phase outflow in the north-west system revealed by the SDSS integrated spectrum, we analysed the publicly available MaNGA data, which cover almost the entire merging system, to study the kinematic and physical properties of cool and warm gas in detail. 
}
{
Galaxy-wide outflowing gas in multiple phases is revealed for the first time in the two merging galaxies. We also detect spatially resolved resonant Na {\small ID} emission associated with the outflows. The derived outflow energetics (mass rate, and kinetic and momentum power) may be consistent with a scenario in which both winds are accelerated by stellar processes and AGN activity, although we favour an AGN origin given the high outflow velocities and the ionisation conditions observed in the outflow regions. Further deeper multi-wavelength observations are required, however, to better constrain the nature of these multi-phase outflows. 
Outflow energetics in the North-West system are strongly different between the ionised and atomic gas components, the latter of which is associated with mass outflow rate and kinetic and momentum powers that are one or two dex higher; those associated with the south-east galaxy are instead similar. 
}
{
Strong kiloparsec-scale outflows are revealed in an ongoing merger system, suggesting that feedback can potentially impact the host galaxy even in the early merger phases. The characterisation of the neutral and ionised gas phases has proved to be crucial for a comprehensive study of the outflow phenomena. 
}

\maketitle

\section[Introduction]{Introduction}

\begin{figure*}[t]
\centering
\includegraphics[width=18.cm,trim=0 0 0 0,clip]{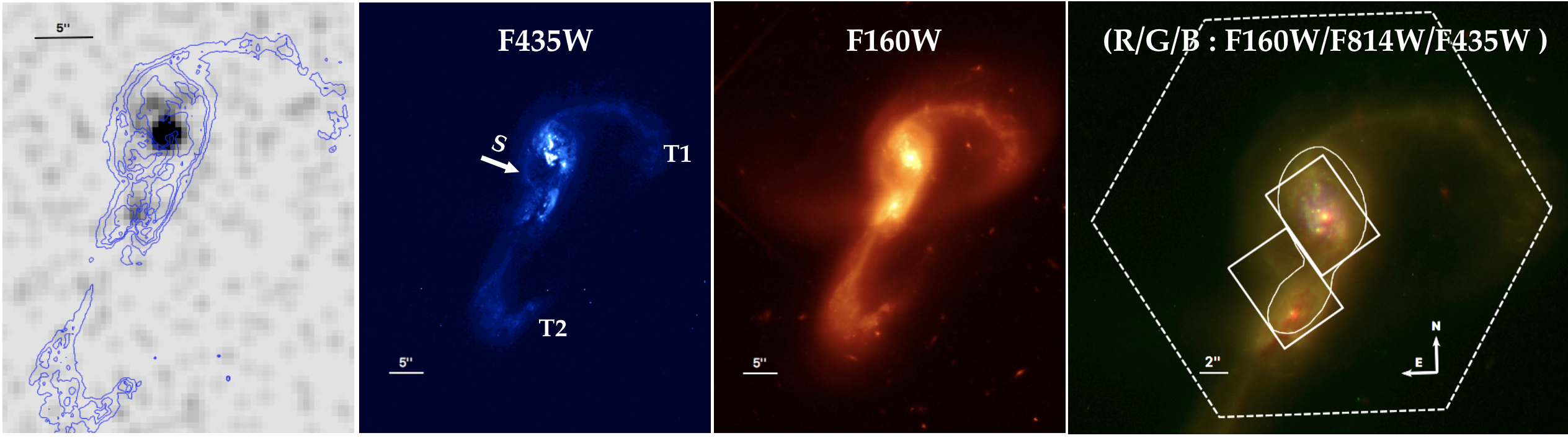}

\caption{\small 
Chandra 0.5-8 keV cutout (left panel) and HST images with colours in log scale. In the Chandra image, blue contours represent F435W fluxes.  In the F435W image, we labelled the two major tidal tails, T1 and T2, that extend over tens of kpc and possibly overlap on the right of the $\mathcal{SE}$ nucleus in the figure. A shell-like region (S) is also indicated.  In the three-colour cutout, the white boxes show the GMOS FOV from which previous spatially resolved kinematic results have been obtained by \cite{Rupke2013}. The white contour represents the S/N = 40 level in the stellar continuum map at $\sim 5550\AA$ and is taken as a reference in the next figures. The MaNGA field of view is also shown with white dashed lines.
}

\label{HSTimages}
\end{figure*}

Powerful active galactic nuclei (AGN) and starburst (SB) -driven outflows are routinely invoked to link the growth of supermassive black holes (SMBH) and galaxies: these phenomena, which deposit energy and momentum into the surrounding medium, are expected to affect the physical and dynamical conditions of infalling gas and thus regulate the formation of new stars and the accretion onto the SMBH (see \citealt{Somerville2015} for a recent review). In particular, feedback phenomena are thought play an important role in the so-called starburst-quasar (SB-QSO) evolutionary sequence framework (e.g. \citealt{Sanders1988}), where a dusty, SB-dominated system, arising from a merger, evolves into an optically luminous QSO (e.g. \citealt{Hopkins2008, Menci2008}). 

Most star-forming galaxies and AGN show evidence of powerful galaxy-wide outflows. Spatially resolved optical, infrared (IR), and millimenter (mm) spectroscopic studies are revealing the extension, morphology, and energetics related to the ejected multi-phase material, in both the local Universe (e.g. \citealt{Arribas2014,Contursi2013,Feruglio2015,Harrison2014,Revalski2018,Sturm2011,Venturi2018b}) and at high redshift (e.g. \citealt{Brusa2016,Brusa2017,Carniani2016,Carniani2017,Cresci2015,Schreiber2014,Genzel2014,Harrison2012,Harrison2016,Newman2012,Perna2015a,Perna2015b,Perna2018,Vietri2018}).
The study of the different phases of the outflowing gas in individual galaxies, which is essential to constrain feedback model prescriptions (see e.g. \citealt{Cicone2018,Harrison2018}),  is limited to a very small number of targets, however  (e.g. \citealt{Brusa2015b,Brusa2017, Carniani2017,Feruglio2015,Feruglio2017, Tombesi2015,Veilleux2017}). 

While most of the evidence of AGN outflows in nearby galaxies comes from the detection of blue-shifted (and red-shifted) outflow components of ionised and molecular gas (generally traced by [O {\small III}]$\lambda 5007$ and CO emission lines), several studies have indicated that only $\sim 1\%$ of nearby AGN show evidence of neutral outflows, which is usually traced by resonant sodium lines at 5890 and 5896$\AA$ (\citealt{Perna2017a} and references therein). Neutral outflows are instead easily detected in ultra-luminous infrared galaxies (ULIRGs) and sources exhibiting concomitant star formation (SF) and AGN activities (\citealt{Nedelchev2017,Cazzoli2016,Rupke2005a,Rupke2017,Sarzi2016}). These findings could suggest that AGN activity has no impact on the neutral gas (e.g. \citealt{Bae2018,Concas2017}), or alternatively, that neutral outflows can be observed only in  obscured AGN in which large reservoirs of cold  gas are still present and the outflow processes can affect all the gas phases (\citealt{Perna2017a}). 

In order to distinguish between the two different scenarios, high-quality spatially resolved spectroscopic observations are required. These early results, which are generally based on spatially integrated data, could have been affected by technical and observational limitations: i) spectra with high signal-to-noise ratios (S/N) are required to trace Na {\small ID} kinematics, ii) a detailed analysis is needed to distinguish between contributions from the  interstellar medium (ISM) and stars to the sodium lines, iii) AGN continuum emission can easily over-shine the Na {\small ID} features, and iv) weak neutral outflows can remain undetected in spatially integrated spectra. All these issues could suggest that the incidence of AGN-driven neutral outflows has been so far underestimated, as also indicated in \citet{Rupke2013,Rupke2015} and \citet{Rupke2017}. The ubiquitous presence of molecular outflows in AGN systems (e.g. \citealt{Fiore2017} for a recent compilation) indeed indicates that AGN winds can successfully accelerate the cold (T $= 10^2-10^3$ K) ISM components. 

\citet{Perna2017a,Perna2017b} presented the spectroscopic analysis of 3$''$ integrated Sloan Digital Sky Survey (SDSS) spectra of $\sim 650$ X-ray detected AGN. We derived the incidence of ionised outflows ($\sim 40\%$) and studied the relations between outflow velocity and different AGN power tracers (SMBH mass, [O {\small III}]$\lambda$5007, and X-ray luminosity). We also derived the incidence of neutral outflows by studying the sodium Na {\small ID}$\lambda\lambda$5890,5896 transitions. In particular, the north-west galaxy in the Mkn 848 system, a binary merger at z $\sim 0.04$, is the only source for which the concomitant presence of outflowing gas in the ionised ([O {\small III}]$\lambda$5007) and neutral phases has been revealed. Our finding is therefore consistent with previous works, which showed a neutral outflow incidence $\lesssim 1\%$ in spatially integrated spectra of local AGN.

Mkn 848  represents an ideal laboratory for feedback studies, as simulations show that gas-rich galaxy mergers can provide the conditions for intense SF and AGN activity, which is essential for the production of powerful outflows. 
The two merging galaxies are above the so-called main sequence (MS) of normal star-forming galaxies (e.g. \citealt{Whitaker2012}), with star formation rates (SFR) of $\sim 80$ M$_\odot$/yr and  $\sim 10$ M$_\odot$/yr for the north-west and south-east galaxies, respectively (\citealt{Yuan2017}), to be compared with an SFR$_{MS} \sim 2$ M$_\odot$/yr of local galaxies with similar stellar masses. Moreover, their optical spectra are associated with starburst-AGN composite emission lines (e.g. \citealt{Yuan2010}, using standard Baldwin-Phillips-Terlevich (BPT) diagnostics; \citealt{Baldwin1981}). 
However, such line ratios, when associated with ULIRGs or interacting galaxies, might also be due to extended shocks, as suggested by \citet{Rich2015}.
Even if the presence of active SMBHs cannot be definitely confirmed by simple optical line ratio diagnostics, Mkn 848 still represents an optimal target for studying the SB-QSO paradigm, as it allows us to test the extreme conditions of the ISM during a merging event.

In this paper, we present a kinematic and physical analysis of Mkn 848 obtained from MaNGA (Mapping Nearby Galaxies at APO) observations, which have been publicly released as part of the SDSS-IV Data Release (\citealt{Blanton2017}).  The manuscript is organised as follows.
In Sect. 2 we present the ancillary target data and discuss the properties of Mkn 848 that we derive from available multi-wavelength information. In Sect. 3 we show the MaNGA data analysis, while Sect. 4 displays the spatially resolved spectroscopic results. In Sect. 5 and Sect. 6 we estimate extinction and hydrogen column density. In Sect. 7 we derive the plasma properties related to systemic and outflowing gas, while in Sect. 8 we investigate the dominant ionisation source for the emitting gas. Finally, we present the outflow energetics in Sect. 9 and summarise our results in the Conclusion.
We adopt the cosmological parameters $H_0 =$ 70 km/s/Mpc, $\Omega_m$ = 0.3, and $\Omega_\Lambda =$ 0.7. The spatial scale of $1''$ corresponds to a physical scale of $\sim 0.8$ kpc at the redshift of Mkn 848.

\section{Ancillary data for Mkn 848}\label{ancillary}
The source Mkn 848 consists of two merging galaxies with similar stellar masses (log(M$_*$/M$_\odot$) $\approx 10.4$; \citealt{Yuan2017}). In Fig. \ref{HSTimages} we show the optical, near-infrared, and three-colour images of the system from the \textit{Hubble Space Telescope} (\textit{HST}), as well as a Chandra X-ray cutout. The two nuclei have a projected separation of 7.5 kpc. Two long, highly curved tidal tails of gas and stars emerge from the north-west ($\mathcal{ NW}$) and south-east ($\mathcal{SE}$) galaxies. One arm (T1 in Fig. \ref{HSTimages}), curving clockwise, stretches from the top of the image and interlocks with the other arm (T2), which curves up counter-clockwise from below. The gravitational interaction between the galaxies may also be responsible for the shell-like region in between the two nuclei (S in the figure). The three-colour image also shows an extended dust lane in the $\mathcal{SE}$ galaxy and strong absorption in the $\mathcal{ NW}$  and $\mathcal{SE}$ nuclear regions.

Available multi-wavelength information from the UV to the IR has recently been analysed by \citet{Yuan2017}: they used photometric and spectroscopic (MaNGA) data to examine the dust attenuation and the star formation activity in six different regions that cover the nuclear cores and tidal arms. The analysis revealed a very recent ($\sim 100$ Myr) interaction-induced starburst in the central regions (especially in the $\mathcal{ NW}$ nucleus), but also signs of earlier ($\sim 500$ Myr) star formation activity in the tidal area. The authors also found that the AGN fraction to the Mkn 848 spectral energy density (SED) is lower than 10\% (see also \citealt{Rupke2013}).  

X-ray emission has been studied for instance by \citet{Iwasawa2011} and \citet{Brightman2011a,Brightman2011b}. Chandra data reveal X-ray emission from the two galaxies, and the $\mathcal{ NW}$ nucleus is much brighter (0.9 dex) than the $\mathcal{ SE}$ nucleus (\citealt{Iwasawa2011}). The two systems are instead not resolved in the (deeper) XMM-Newton observations (\citealt{Brightman2011a}), from which more detailed information can be derived for the $\mathcal{ NW}$ system (assuming that most of the X-ray flux comes from the $\mathcal{ NW}$ regions; \citealt{Iwasawa2011}). From the XMM-Newton spectroscopic analysis we derive an observed  L$_{0.5-2\ keV} =1.6\times 10^{41}$ erg/s; this soft-band emission can be entirely explained by stellar processes (for an SFR $= 80$ M$_\odot$/yr we can derive an expected L$_{0.5-2\ keV} \approx 2\times 10^{41}$/erg/s; \citealt{Mineo2013}).
Instead, the hard-band shows an excess emission (observed L$_{2-10\ keV} =3\times 10^{41}$ erg/s with respect to $\approx 10^{41}$ erg/s expected from the SFR, \citealt{Lehmer2010}, although the two values are statistically consistent). The XMM hard spectrum is better reproduced with a heavily obscured AGN, with N$_H =1.4\times 10^{24}$ cm$^2$ and intrinsic luminosity L$_{2-10} = 3\times 10^{42}$ erg/s; the lower confidence limits (c.l.) are of the order of 0.4 dex, while the upper c.l. are unconstrained by the data, which might indicate even higher column densities and intrinsic luminosities. A bolometric luminosity of L$_{bol} =3\times 10^{43}$ erg/s can be inferred for the $\mathcal{ NW}$  SMBH by applying an X-ray bolometric correction of 10 (\citealt{Lusso2012}). With the currently available archival Chandra and XMM observations, we are unable to derive stringent constraints on the $\mathcal{ SE}$ X-ray (and bolometric) luminosity and cannot distinguish between dominating AGN and stellar activity in this system. Future observations will allow us to better explore the potentially dual AGN nature of Mkn 848 through hard X-ray emission.

The nuclear optical emission has been studied in detail by \citet{Rupke2013}: they observed the $\mathcal{NW}$ and $\mathcal{SE}$ nuclear regions (white boxes in Fig. \ref{HSTimages}, right panel), with the Integral Field Unit (IFU) GMOS at the Gemini North telescope.  With total exposure times of 2h ($\mathcal{SE}$) and 2h 30min ($\mathcal{NW}$), they detected high-$v$, blue-shifted ionised (H$\alpha$ and [N {\small II}]$\lambda\lambda$6549,6585) and atomic (Na {\small ID}) gas associated with conical outflow in the $\mathcal{NW}$ system, and the possible presence of ionised outflow in the $\mathcal{SE}$ galaxy.
Both nuclei were classified as SB-AGN composite systems, and the observed outflows were associated with SF-driven winds.
The availability of the deeper MaNGA observations, which have a   larger field of view (FOV; dashed line in Fig. \ref{HSTimages}, right panel) and larger wavelength range than the data presented by \citet{Rupke2013}, allowed us to perform a more detailed analysis of the system. 

We also note that archival Mkn 848 mm observations could also suggest the presence of molecular outflows. Carbon monoxide CO(3-2) line observations have been obtained  by \citet{Leech2010} with the James Clerk Maxwell Telescope. These observations covered the two nuclear regions with a single-pointing and revealed a well-detected, narrow CO profile, with an FWHM $\sim 120$ km/s, and the possible presence of faint extended wings at both negative and positive velocities of up to $\approx \pm 250$ km/s. CO(1-0) and CO(2-1) observations have previously also been reported (\citealt{Sanders1991,Chini1992}, respectively), but their lower S/N does not permit to confirm or rule out the presence of such high-$v$ wings (see also \citealt{Yao2003}). 

These are some reasons why Mkn 848 can be considered as an ideal laboratory for the study of the SB-QSO galaxy evolution paradigm. In the next sections we present the MaNGA spectroscopic analysis we performed to fully characterise the dynamical and physical conditions in this system.

\section{Spectroscopic analysis of the MaNGA data}\label{spectroscopicanalysis}

MaNGA is a new SDSS survey designed to map the composition and kinematic structure of 10$^4$ nearby galaxies using IFU spectroscopy. Its wavelength coverage extends from 3600 to 10400$\AA$, with a spectral resolution of $\sim 2000$. The diameters of MaNGA data-cubes range from 12$''$ to 32$''$, with a spatial sampling of 19 to 127 fibers. Mkn 848 (MaNGA ID 12-193481) has been targeted with 127 fibers, which almost entirely covers the merging system. The spectroscopic analysis presented in this work takes advantage of the automatic fully reduced MaNGA data release (we refer to \citealt{Law2016} for the MaNGA data reduction pipeline).
 
We performed the data-cube spectral fit following the prescriptions in our previous studies (e.g. \citealt{Perna2017a}) and adapted them here to trace the kinematic properties of the neutral and ionised ISM components of Mkn 848. 
Prior to modelling the ISM features, we used pPXF routines (\citealt{Cappellari2004,Cappellari2016}) to subtract the stellar contribution from continuum emission and absorption line systems, using the MILES stellar population models (\citealt{Vazdekis2010}).  The separation between stellar and ISM contributions is required in particular to properly measure the kinematics of neutral gas in the ISM. The pPXF fit was performed on binned spaxels using a Voronoi tessellation (\citealt{Cappellari2003}) to achieve a minimum S/N $> 50$ on the continuum in the [$5380 \div 5770$]$\AA$ rest-frame wavelength range, modelling the stellar features in the wavelength range 3550-7400$\AA$. 
A pure emission and absorption ISM line data-cube was therefore created by subtracting the best-fitting pPXF models.

\subsection{Simultaneous modelling of optical ISM features}

We used the Levenberg–Markwardt least-squares fitting code MPFITFUN (Markwardt 2009) to simultaneously reproduce the ISM line data-cube with Gaussian profiles, that is, the H$\beta$, [O {\small III}]$\lambda\lambda$4959,5007 doublet and He {\small I}$\lambda$5876 emission lines, the Na {\small ID} absorption line doublet at $\lambda\lambda$5891,5897, and the [O {\small I}]$\lambda$6300, H$\alpha$, [N {\small II}]$\lambda\lambda$6549,6585, and [S {\small II}]$\lambda\lambda$6718,6732 emission lines. 

As mentioned above, atomic and ionised outflowing material has been observed in the $\mathcal{NW}$ component of Mkn 848: prominent blue wings have been well detected in the rest-frame optical emission lines and in Na {\small ID} absorption features, tracing high-velocity approaching warm and cold material, with $\sim -900$ km/s and $\sim -1200$ km/s, respectively. We therefore modelled these line features as described below.  

\subsubsection{Ionised lines}
We used three (at maximum) sets of Gaussian profiles to reproduce the asymmetric ionised emission line features:
\begin{itemize}
\item[NC]
-- One narrow component (NC) set: 10 Gaussian lines, one for each emission line (H$\beta$, the [O {\small III}] doublet, He {\small I},  [O {\small I}], H$\alpha$, and the [N {\small II}] and [S {\small II}] doublets) to account for unperturbed systemic emission. The width (FWHM) of this kinematic component was set to be $\lesssim 400$ km/s.
\item[OC]
-- One (two) outflow component (OC) set(s): nine Gaussian lines, one for each emission line (without He {\small I}, which is usually faint and never well constrained) to account for outflowing ionised gas. No upper limits were fixed for the width of this kinematic component.  
\end{itemize} 

We constrained the wavelength separation between emission lines within a given set of Gaussian profiles using the rest-frame vacuum wavelengths. 
Moreover, the relative flux of the two [N {\small II}] and [O {\small III}] doublet components was fixed to 2.99 and the [S {\small II}] flux ratio was required to be within the range $0.44<f(\lambda6716)/f(\lambda6731)<1.42$ (\citealt{Osterbrock2006}). All emission line features were fitted simultaneously to significantly reduce the number of free parameters and thus the model degeneracy. 

\subsubsection{Neutral lines}

The sodium absorption lines were fitted simultaneously to the emission features described above. However, a different approach was required to model these lines. Na {\small ID}  is a resonant system, and both absorption and emission contributions can be observed across the extension of a galaxy (e.g. \citealt{Rupke2015,Prochaska2011}). 
We note that sodium emission is not a common feature: it has been observed at very low level in stacked SDSS spectra (\citealt{Chen2010, Concas2017}) and in a few nearby galaxies (see \citealt{Rupke2015} and references therein). Unconstrained in the previous observations, it is clearly detected in the MaNGA data-cube (see Fig. \ref{NaIDspectra}). 

The modelling of sodium resonant transitions is made all the more difficult because the Na {\small  ID} system is generally faint, and both systemic and outflowing gas can be associated with emission and absorption contributions (e.g. \citealt{Prochaska2011}). 
To reproduce the sodium absorption lines, three (at maximum) sets of Gaussians were used:
\begin{itemize}
\item[NC$^\pm$]
-- One narrow component: two Gaussian lines, one for each Na {\small ID} transition, to account for (emitting or absorbing) systemic gas. Gaussian amplitudes can vary from negative to positive values to account for absorbing or emitting unperturbed gas. Their kinematic properties (i.e. line widths and centroids) were set to be equal to those of the NC emission lines.
\item[OC$^-_1$]
-- One outflow component for absorbing gas: two Gaussian lines, one for each emission line of the doublet, to account for outflowing absorbing neutral gas. No upper limits were fixed for the width of this kinematic component, and no constraints were set for the Gaussian centroids (but their relative separation was fixed).
\item[OC$^-_2$]
-- A second outflow component for absorbing gas: two Gaussian profiles, to account for the presence of very high velocity outflowing gas. Their line widths were fixed to values higher than 700 km/s; no constraints were set for the Gaussian centroids (but, as before, their relative separation was fixed). 
\item[OC$^+$]
-- One outflow component for emitting gas: one Gaussian profile, to account for the possible presence of faint wings in emitting Na {\small ID} profile. No constraints were set for this Gaussian profile. In our procedure, this component was always fitted together with the sodium OC$^-$ Gaussian profiles.
\end{itemize} 

\begin{figure*}[h]
\centering
\includegraphics[width=9.cm,trim=15 0 60 0,clip]{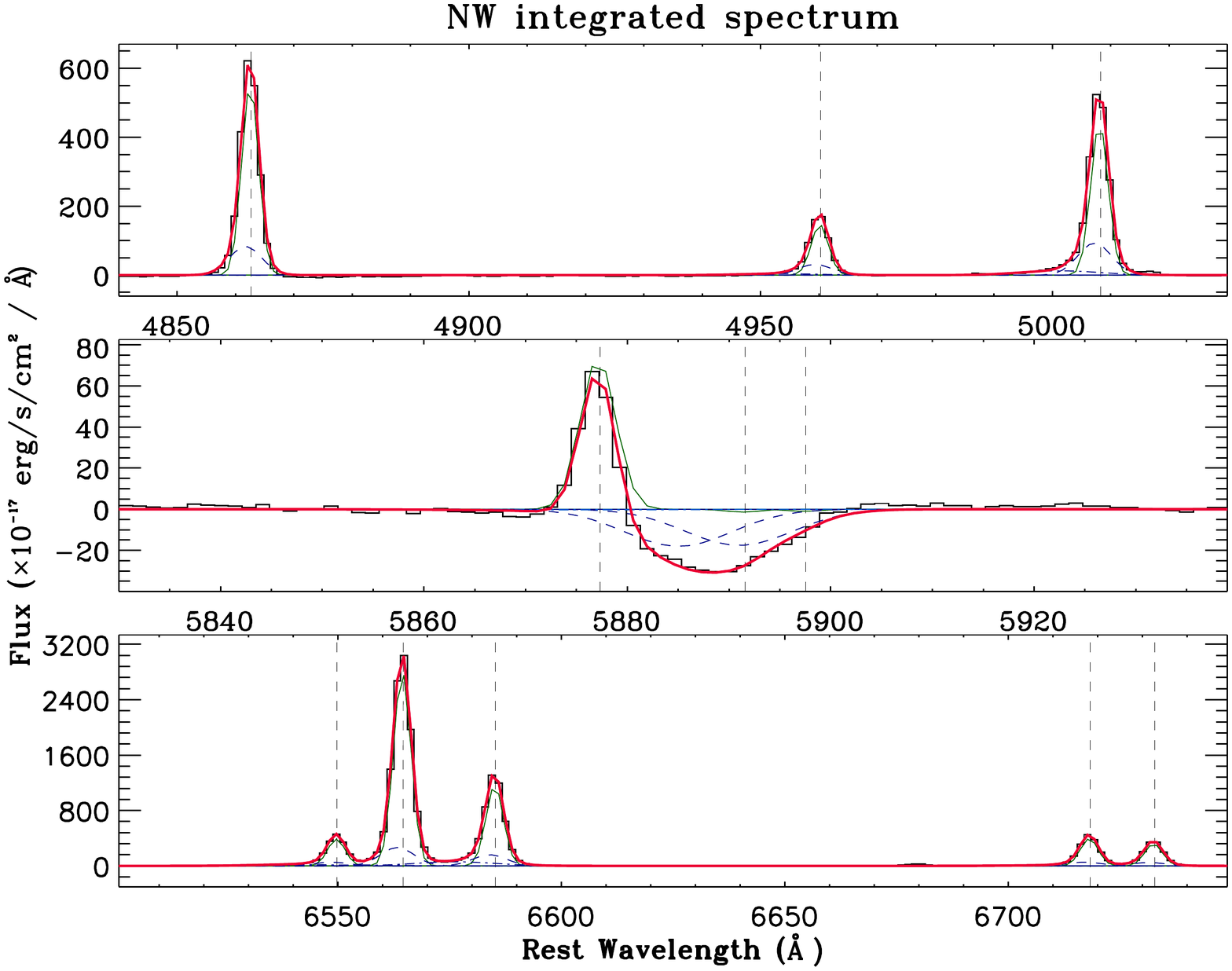}
\hspace{0.01cm}
\includegraphics[width=9.cm,trim=15 0 60 0,clip]{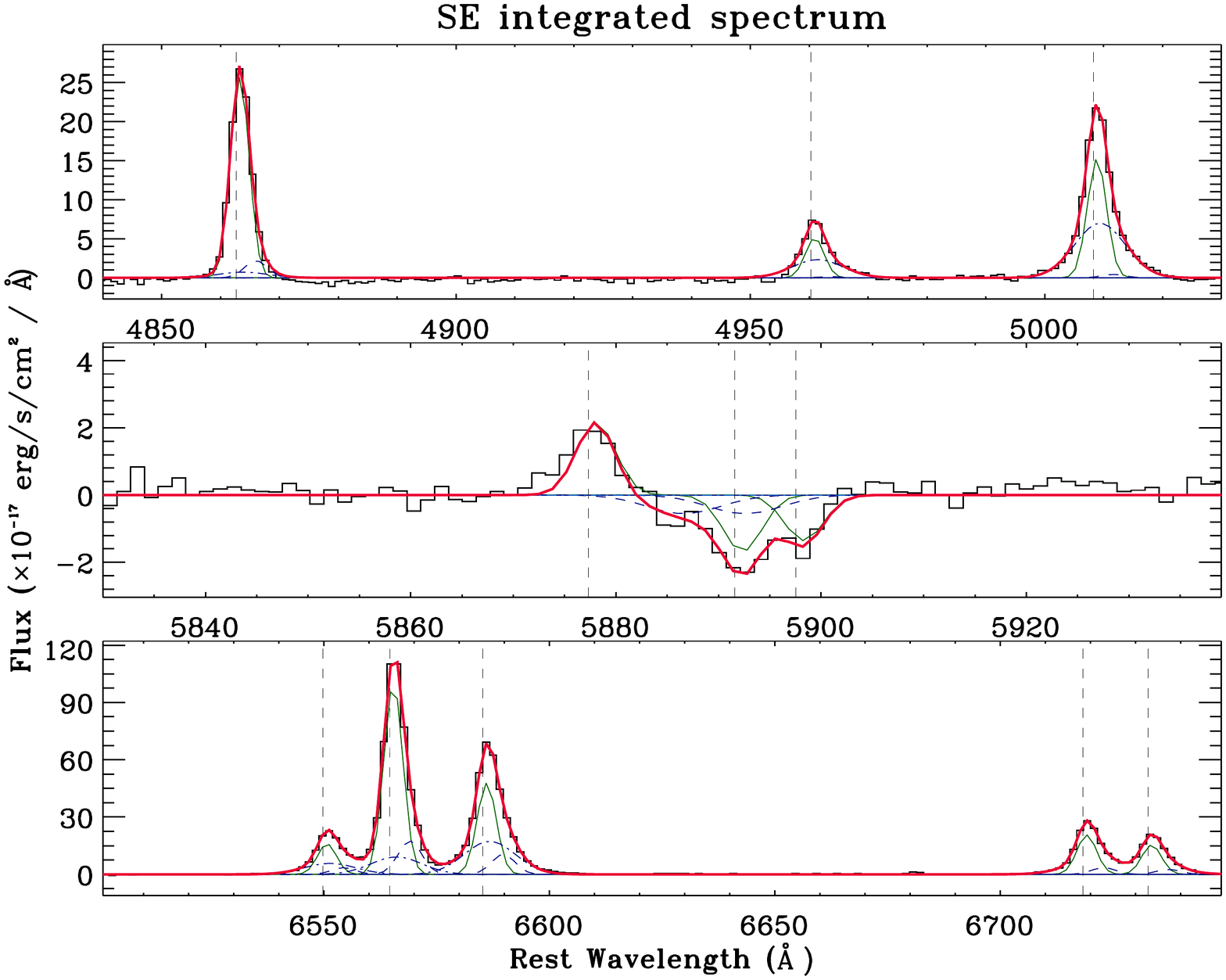}

\caption{\small 
 Portion of MANGA integrated spectra of Mkn 848, for the $\mathcal{NW}$ and $\mathcal{SE}$ regions, with superimposed best-fitting components. {\it Top panels:} [O III]+H$\beta$ emission lines from the $\mathcal{NW}$ ({\it left panel}) and $\mathcal{SE}$ ({\it right}) regions. {\it Central panels:} He I emission line and Na ID absorption line system. 
{\it Bottom panels:} [N II], H$\alpha$ and [S II] lines. The different sets of Gaussian components are shown with solid green curves (NC), dashed and dot-dashed blue lines (OC). Stellar continuum emission has been previously modelled with the pPXF routine and subtracted. The dotted lines mark the wavelengths of H$\beta$ and [O III]$\lambda\lambda$4959,5007, in the top panels, He I and Na ID doublet lines in the central panels, and [N II]$\lambda$6548, H$\alpha$, [N II]$\lambda$6581 and [S II] doublet lines in the bottom panels. The systemic redshift is defined on the basis of the best-fit solution for the NC in the $\mathcal{NW}$ integrated spectrum. The $\mathcal{SE}$ systemic lines show a velocity offset of $\sim 90$ km/s. The best-fits reproduce all but the Na ID NC in the $\mathcal{NW}$ integrated spectrum with high significance. 
}

\label{integratedspectra}
\end{figure*}

The reduced number of Gaussian profiles in the OC$^+$ (with respect to the absorption) is justified by the necessity of minimising the degeneracy in best-fit models; in addition, we note that this broad emission component is generally very faint (see e.g. Fig. \ref{NaIDspectra}). 
Instead, the two Gaussian profiles used for the negative OCs are generally associated with stronger components (Fig. \ref{integratedspectra}). 

Because the Na {\small ID} system consists of two transitions of the same ion, the flux ratio of the $f$($\lambda5890$) and $f$($\lambda5896$) lines (hereinafter referred to as the H and K features, respectively) can vary between the optically thick ($f(H)/f(K)=1$) and thin ($f(H)/f(K)=2$) limits (e.g. \citealt{Bechtold2003,Rupke2015}); we therefore constrained the flux ratio within the range $1<f(H)/f(K)<2$ for all NC$^\pm$,  OC$^-_1$ , and OC$^-_2$ Gaussians. 

\subsubsection{Bayesian information criterion}
We performed each spectral fit three times at maximum, with one to three Gaussian sets  (e.g. NC, and one or two OC sets for the ionised component) to reproduce complex line profiles where needed.
The number of sets used to model the spectra was derived on the basis of the Bayesian information criterion (BIC; \citealt{Schwarz1978}), which uses differences in $\chi^2$ that penalise models with more free parameters. In more detail, for an individual model we computed the BIC as $\chi^2 + k\ ln(N)$, with $N$ the number of data points and $k$ the number of parameters of the model. For an individual spectrum, we first derived $\Delta BIC_{1-2}=BIC_1-BIC_2$, where $BIC_1$ and $BIC_2$ are the values derived from the models with one and two sets of Gaussians, respectively. When  $\Delta BIC_{1-2}$ was lower than a given threshold ($\Delta BIC_{1-2}<30$; \citealt{Mukherjee1998,Harrison2016}), we favoured the fit with the lower number of components, that is, the one-component fit; otherwise, we performed an additional fit with three components and favoured the two- or three-component models according to the $\Delta BIC_{2-3}$ value, using the threshold described above.

\begin{figure*}[h]
\centering
\includegraphics[width=9.cm,trim=15 0 60 0,clip]{{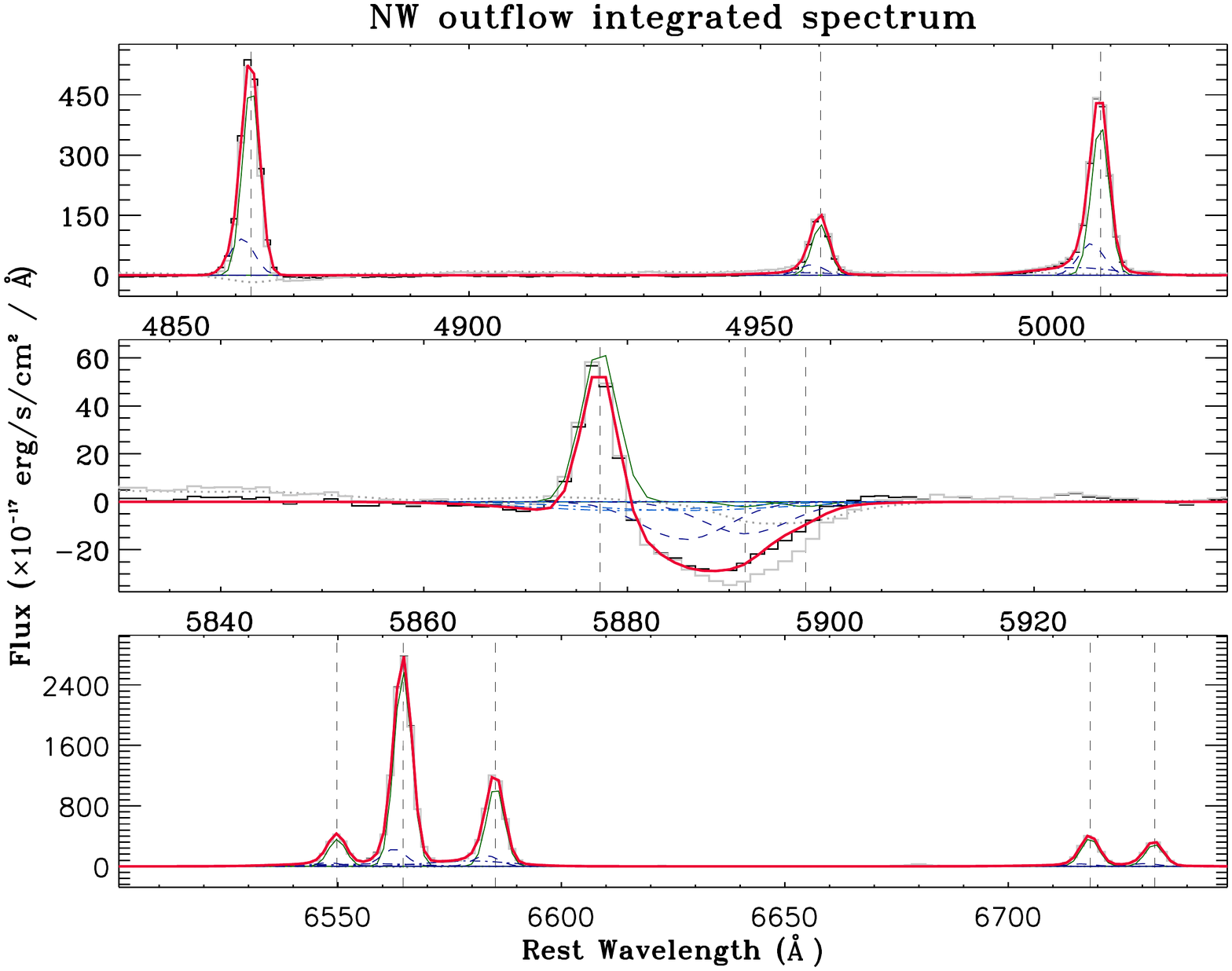}}
\hspace{0.01cm}
\includegraphics[width=9.cm,trim=15 0 60 0,clip]{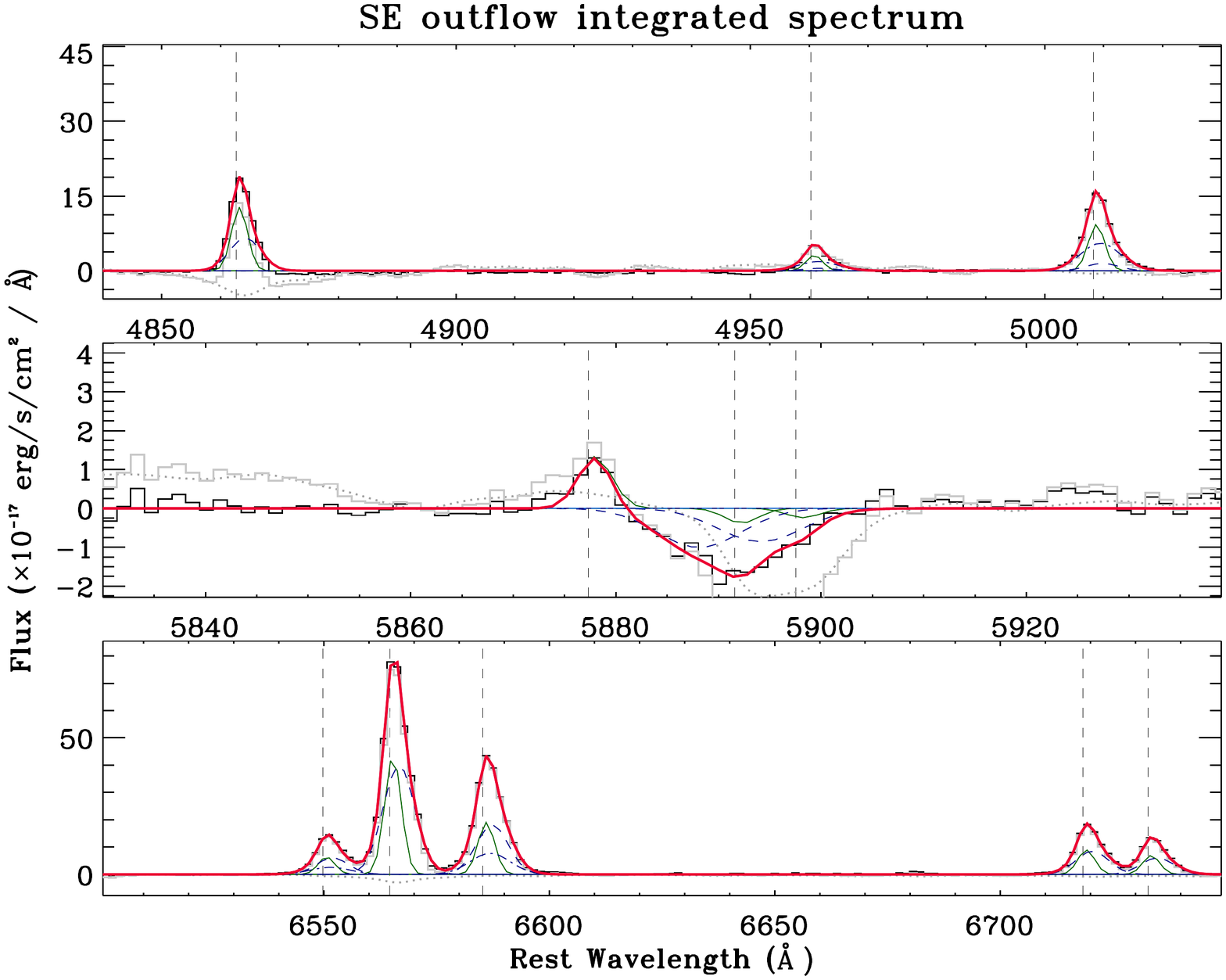}

\caption{\small 
$\mathcal{NW}$ ({\it left}) and $\mathcal{SE}$ ({\it right}) outflow-integrated spectra  extracted from the spaxels associated with high-velocity gas in the neutral and/or ionised phases ($\Delta V_{02}([O III]) < -550$ km/s and/or $\Delta V_{02}(Na ID) <  -900$ km/s , for the $\mathcal{NW}$ system; $\Delta V_{98}([O III]) > 450$ km/s and/or  $\Delta V_{02}(Na ID) <-750$ km/s, for the $\mathcal{SE}$ galaxy; see Sect. \ref{Sspatiallyresolved}). The shell-like region spaxels are not included in these integrated spectra. The stellar continuum emission has been previously modelled and subtracted from each spaxel. 
The red curves overplotted on the spectra represent the best-fit solutions that reproduce the line profiles; NC and OC Gaussians are also shown as in Fig. \ref{integratedspectra}.
The outflow-integrated spectra extracted from the original MaNGA datacube (hence without pPXF best-fit model subtraction) and the  integrated pPXF best-fit spectra are also shown in the figure with solid and dotted grey curves, respectively. 
For the two grey curves we applied a downward shift for visualisation purposes.
}

\label{outflowintegratedspectra}
\end{figure*}
\subsection{Non-parametric velocity analysis} 

The simultaneous line-fitting technique introduced in the previous sections allows us to distinguish between unperturbed gas components (NC), which are associated with emitting gas under the influence of the gravitational potential of the system, and the OC, which most probably is associated with emission of outflowing gas. Therefore, this modelling allows a good separate estimate of the unperturbed and outflowing gas fluxes.
Instead, to characterise the kinematic properties of the ISM gas across the FOV best  in a homogeneous way and to avoid any dependence on the number of distinct Gaussians used to model the line features in individual spaxels, we used the non-parametric approach (e.g. \citealt{Zakamska2014}). Non-parametric velocities were obtained by measuring the velocity $v$ at which a given fraction of the total best-fit line flux is collected using a cumulative function. The zero-point of velocity space was defined by adopting the systemic redshift derived from the $\mathcal{NW}$ integrated spectrum (Sect. \ref{Sintegratedspectra}). We carried out the following velocity measurements on the best fit of [O {\small III}]$\lambda$5007 ([O {\small III}] hereinafter) and H$\alpha$ line profiles for the ionised gas and of the Na {\small ID} to trace the neutral ISM component:

\begin{itemize}
\item
$W80$: The line width comprising 80\% of the flux, which is defined as the difference between the velocity at 90\% ($V_{90}$) and 10\% ($V_{10}$) of the cumulative flux.
\item
$V_{02}$: The maximum blue-shift velocity parameters, which is defined as the velocity at 2\% of the cumulative flux. 
\item
$V_{98}$: The maximum red-shift velocity parameters, which is defined as the velocity at 98\% of the cumulative flux. 
\item
$V_{p}$: The velocity associated with the emission line peak.
\end{itemize}

These velocities are usually introduced to characterise asymmetric line profiles. In particular, $W80$ is closely related to the velocity dispersion for a Gaussian velocity profile, and is therefore indicative of the line widths, while  maximum velocities are mostly used to constrain outflow kinematics when prominent line wings are found (e.g. \citealt{Cresci2015,Perna2017a}; see also e.g. \citealt{Harrison2018}). We used the 2\% and 98\% velocities to facilitate comparison with \cite{Rupke2013}, although the authors defined the maximum blue-shifted velocity ($V_{98}$ in their work) in a slightly different way and derived the outflow velocity from the broad (outflow) component alone. As a result, their estimates may be higher than our values (which are defined from the total emission line profile). Finally, $V_p$, which generally traces the velocity of the brightest narrow emission line components, can been used as a kinematic tracer for rotating-disc emission in the host galaxy (e.g. \citealt{Harrison2014, Brusa2016, Perna2018}).

We note that perturbed kinematics that are due to gravitational interactions of the two galaxies could also be present in the system, especially along the tidal features. We used our multi-component models and the non-parametric techniques to distinguish between gravitational interaction and rotational and outflow signatures.

\section{Cool and warm gas kinematics}
In this section we present the results we obtained from the fitting prescription and the non-parametric approach exposed above. We show integrated spectra with high S/N of the two merging galaxies and the spatially resolved kinematic and physical properties of Mkn 848. 

\begin{figure}[t]
\centering
\includegraphics[width=9.cm,trim=105 50 350 180,clip]{{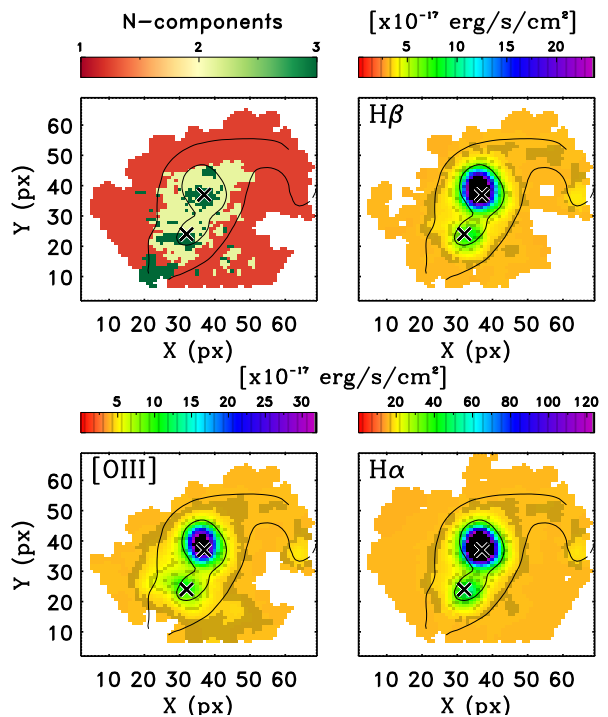}}

\caption{\small 
{\it Top-left panel:} Number of sets of Gaussians we used to model individual spaxel spectra. {\it Top-right and bottom panels:} H$\beta$, [O III], and H$\alpha$ integrated fluxes obtained from simultaneous best fits (we show only the spaxels with an S/N $> 3$).  Solid contours refer to S/N levels of 10 and 40 in the stellar continuum map (at $\sim 5550\AA$). Black crosses indicate the nuclear regions.
}

\label{Ngau}
\end{figure}

\subsection{Integrated spectra}\label{Sintegratedspectra}
Before analysing the spatially resolved information of the ISM in detail, we extracted the integrated $\mathcal{ NW}$ and $\mathcal{SE}$ nuclear spectra from the MaNGA data-cube, considering the spaxels associated with S/N$>45$ in the stellar continuum at $\sim 5550\AA$. The two spectra are shown in Fig. \ref{integratedspectra}. The $\mathcal{NW}$ and $\mathcal{SE}$ galaxies are at z $=$ 0.0404 and 0.0407, respectively. In the following analysis, we derived all velocities using the systemic of the NC in the $\mathcal{NW}$ integrated spectrum as a reference unless stated otherwise.

Asymmetric profiles with extended blue and red wings are observed in all emission lines. In particular, OC Gaussians are blue-shifted with respect to the NC in the $\mathcal{NW}$ region; in the $\mathcal{SE}$ spectrum, all the OCs are instead red-shifted with respect to the narrow Gaussians. The red-shifted emission in the $\mathcal{SE}$ nucleus was not detected by \cite{Rupke2013}, probably because of the low fluxes associated with this nucleus ($f_{SE}:f_{NW}\sim 1:10$ in all the emission line features; Fig. \ref{integratedspectra}). We also note that all the profile peaks are associated with the NC Gaussians, and that outflowing gas therefore does not dominate in the emission of ionised gas.

The sodium absorption systems display very complex profiles. In the $\mathcal{NW}$ spectrum, the NC are faint and poorly constrained (but their kinematic properties are fixed to those of the other emission lines), and the bulk of the absorption is associated with outflowing gas. 
The $\mathcal{SE}$ spectrum displays similar complexities, with prominent outflow components. In this case, the neutral outflow in the $\mathcal{SE}$ system was not previously detected with the GMOS instrument either (\citealt{Rupke2013}).

Taking advantage from the spatially resolved analysis presented in the next sections, we also show in Fig. \ref{outflowintegratedspectra} the outflow-integrated spectra, which were obtained considering the only spaxels associated with very high-velocity in the neutral and/or ionised phases. These spectra highlight the high-velocity blue and red wings in the ionised gas line profiles, and more importantly, the complexities in the emission and absorption line profiles. In particular, the [O {\small III}] and Na {\small ID}  outflow tracers were fitted with two or three Gaussian components. The very broad and smoothed profiles do not allow us to uniquely identify the individual kinematic components associated with the outflows for either the ionised or atomic phases.

In Table \ref{spectralresults} we report the spectroscopic results obtained from modelling the $\mathcal{NW}$ and $\mathcal{SE}$ integrated spectra. In the next sections, we introduce different line ratio diagnostics to constrain the ISM conditions (e.g. dust extinction and ionisation; see Table \ref{spectralresults}). In advance of this, we note that the flux ratios $f([{\rm O { \small III}}])/f(H\beta)$ and $f([{\rm N {\small II}}])/f(H\alpha)$ that are normally used to separate purely SF galaxies from galaxies containing AGN (\citealt{Baldwin1981}) suggest a composite SB-AGN ionisation in the two systems. The outflowing gas is instead associated with higher flux ratios that are compatible with AGN ionisation (see Sect. \ref{ionisation} for details).
\begin{figure*}[p]
\centering
\begin{minipage}{1\linewidth}
\includegraphics[width=18.cm,trim=0 180 0 50,clip]{{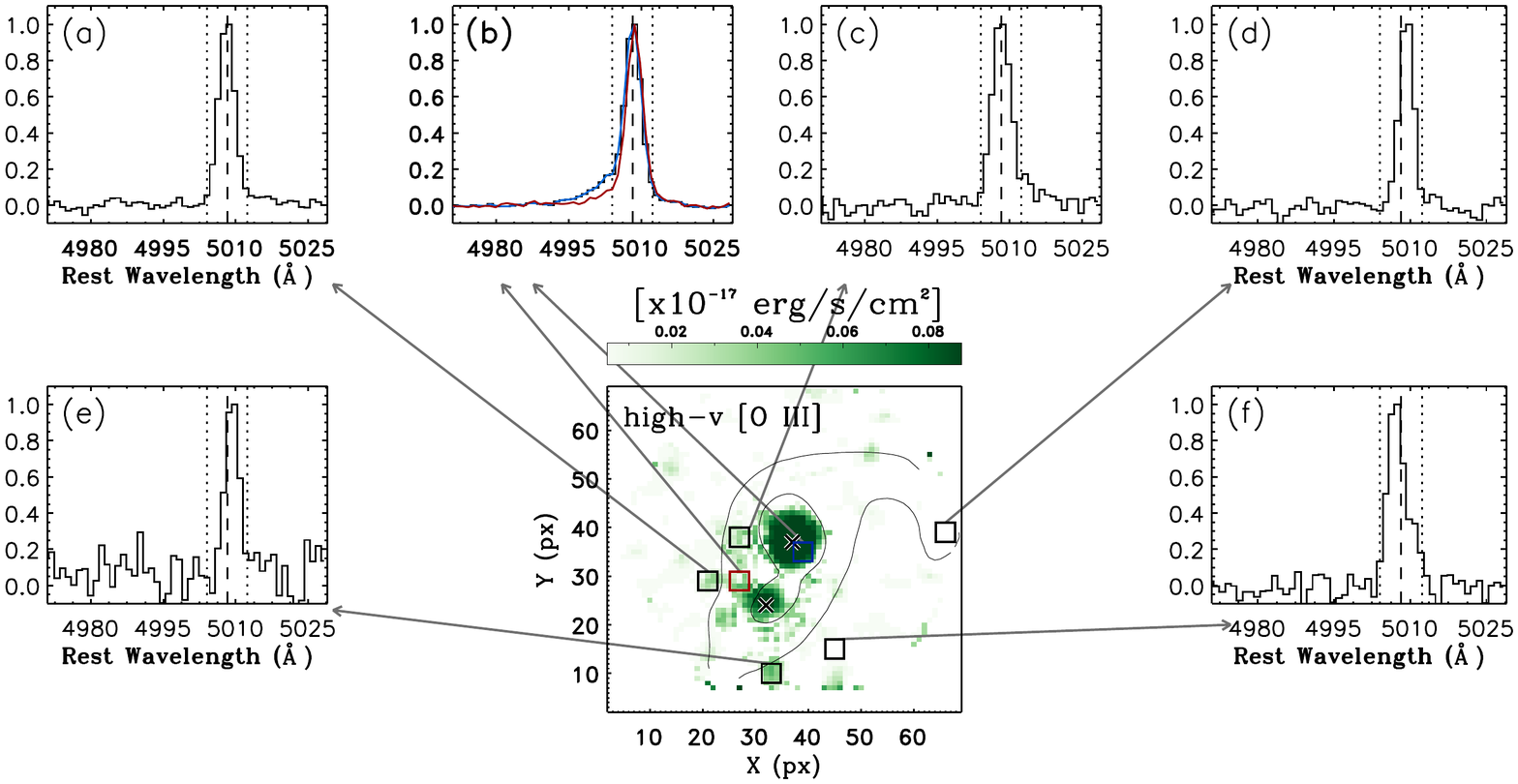}}
\caption{\small 
High-velocity [O III]$\lambda$5007 emission line flux across the FOV obtained by integrating for each spaxel the spectra in the regions $4890-5004\AA$ and $5014-5030\AA$, which correspond to absolute velocities $>250$ km/s with respect to the reference frame of the $\mathcal{NW}$ nucleus. The map was obtained after subtracting the stellar contribution (using pPXF best-fit spaxel spectra). Solid contours refer to the S/N =10 and 40 levels in the stellar continuum map. 
Different spectra, extracted from the regions labelled in the flux maps, are also shown in the insets, with normalised fluxes; blue and red spectra in inset (b) show the most asymmetric line profiles and are extracted from the $3\times 3$ pixel regions shown in the central panel with blue and red squares, respectively. Vertical dashed lines indicate the ($\mathcal{NW}$ nucleus systemic) wavelength of the [O III]$\lambda$5007 line; dotted lines refer to $\pm 250$ km/s with respect to the $\mathcal{NW}$ nucleus systemic velocity. 
}
\label{OIIIspectra}
\end{minipage} \par\medskip
\begin{minipage}{1\linewidth}


\centering
\includegraphics[width=18.cm,trim=0 0 0 180,clip,angle=180]{{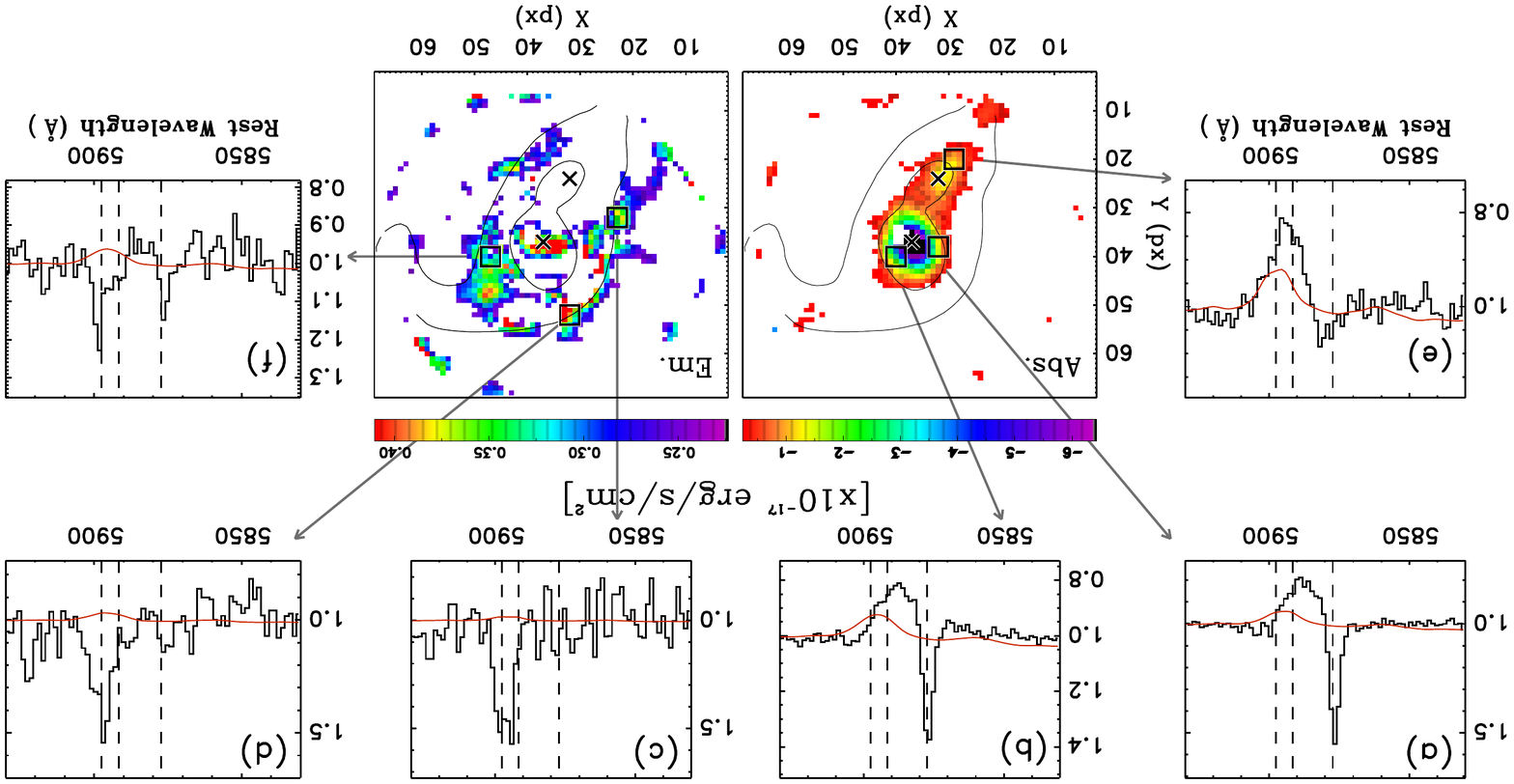}}
\caption{\small 
{\it Central panels:} Na ID absorption (Abs.) and emission (Em.) line fluxes across the FOV obtained by integrating for each spaxel the spectra in the regions $5860-5910\AA$ and $5890-5920\AA$, respectively. Negative and positive fluxes were obtained after subtracting the stellar contribution (pPXF best fit); for the absorption fluxes, we also removed the He I emission line (simultaneous best-fit results). {\it Insets:} Different spectra extracted from the regions labelled in the flux maps, with normalised fluxes: absorption- (emission-) dominated Na ID line profiles are shown in the right (left) panels. Red curves represent the Na ID stellar contribution fitted with pPXF, rescaled for visual purposes. Vertical dashed lines indicate the ($\mathcal{NW}$ nucleus systemic) wavelengths of the He I and Na ID doublet lines. 
}
\label{NaIDspectra}
\end{minipage} \par\medskip
\end{figure*}

\begin{table*}
\footnotesize
\begin{minipage}[!h]{1\linewidth}
\setlength{\tabcolsep}{3.1pt}
\centering
\caption{Spectral fit results}
\begin{tabular}{lc|cccccc|ccc}
\multicolumn{11}{c}{Nucleus-integrated spectra}\\
   && $f([O III])$ & $f([O III])/H\beta$ & $f(H\alpha)$ & $f([N II])/f(H\alpha)$&$f([S II])/f(H\alpha)$& $A_V$ & $V_{02}([O III])$& $V_{02}(Na ID)$ & $V_{98}([O III])$\\
\toprule
$\mathcal{NW}$ &tot       &$15425\pm 208$ & $0.98\pm 0.01$&$56040\pm 587$&$0.52\pm 0.01$&$0.30\pm 0.01$& $1.90\pm 0.01$ &$-742\pm 32$& $-687\pm 23$ & $235\pm 3$\\
  &NC       &$11628\pm 242$ & $0.81\pm 0.01$&$52058\pm 741$&$0.40\pm 0.01$& $0.24\pm 0.01$&$2.14\pm 0.02$ & & \\
  &OC       &$2980\pm 274$ & $1.43\pm 0.05$ & $6998\pm 588$&$0.97\pm 0.03$& $0.44\pm 0.03$&$1.13\pm 0.09$ && & \\
\hline
$\mathcal{SE}$ &tot         & $914\pm 29$ & $1.18\pm 0.02$&$2770 \pm 58$ & $0.74\pm 0.01$ &$0.48\pm 0.01$&$1.98\pm 0.03$& $-403\pm 10$ & $-340\pm 27$ & $514\pm 9$ \\
  & NC       & $136 \pm 16$ & $0.70\pm 0.03$& $649 \pm 81$ & $0.47\pm 0.01$ &$0.37\pm 0.06$&$1.12\pm 0.08$& & & \\
  &OC        & $1328 \pm 155$ & $1.80\pm 0.12$& $2789 \pm 225$ & $0.92\pm 0.03$ &$0.56\pm 0.06$&$2.80\pm 0.13$& & & \\
\toprule
\\
\multicolumn{11}{c}{Outflow-integrated spectra}\\
   && $f([O III])$ & $f([O III])/H\beta$ & $f(H\alpha)$ & $f([N II])/f(H\alpha)$&$f([S II])/f(H\alpha)$& $A_V$ & $V_{02}([O III])$& $V_{02}(Na ID)$ & $V_{98}([O III])$\\
\toprule
$\mathcal{NW}$ &tot       &$15800\pm 230$ & $0.98\pm 0.01$ &$57700\pm 600$&$0.52\pm 0.01$&$0.28\pm 0.01$& $2.08\pm 0.01$ &$-740\pm 20$& $-730\pm 18$ & $250\pm 9$\\
                          &NC       &$12300\pm 190$ & $0.82\pm 0.01$ &$54700\pm 600$&$0.41\pm 0.01$& $0.25\pm 0.02$&$2.25\pm 0.04$ & & \\
                          &OC       &$2540\pm 230$   & $1.64\pm 0.11$ & $5270\pm 450$ &$1.15\pm 0.06$& $0.46\pm 0.04$&$1.29\pm 0.11$ && & \\
\hline
$\mathcal{SE}$ &tot         & $544\pm 18$     & $1.02\pm 0.02$&$1900 \pm 42$    & $0.60\pm 0.01$ &$0.41\pm 0.01$&$1.91\pm 0.03$& $-317\pm 12$ & $-460\pm 12$ & $450\pm 11$ \\
                         & NC       & $93 \pm 10$      & $0.76\pm 0.05$& $410 \pm 30$     & $0.44\pm 0.01$ &$0.37\pm 0.06$&$1.07\pm 0.09$& & & \\
                         &OC        & $634 \pm 80$    & $1.29\pm 0.09$& $1820 \pm 150$ & $0.70\pm 0.03$ &$0.44\pm 0.07$&$2.56\pm 0.13$& & & \\
\toprule
\end{tabular}
\label{spectralresults}
\end{minipage}
Note: Emission line fluxes are in units of $10^{-17}$ erg/s/cm$^2$ and are corrected for extinction, using the visual extinction ($A_V$) values derived from Balmer decrement ratios. Non-parametric velocities are derived here using different zero-points in the velocity space, assuming  redshifts of  0.404 and 0.407 for the $\mathcal{NW}$ and $\mathcal{SE}$ systems, respectively.  
\end{table*}

\subsection{Spatially resolved spectroscopy}\label{Sspatiallyresolved}

\begin{figure*}[h]
\centering
\includegraphics[width=13.cm,trim=250 410 99 30,clip]{{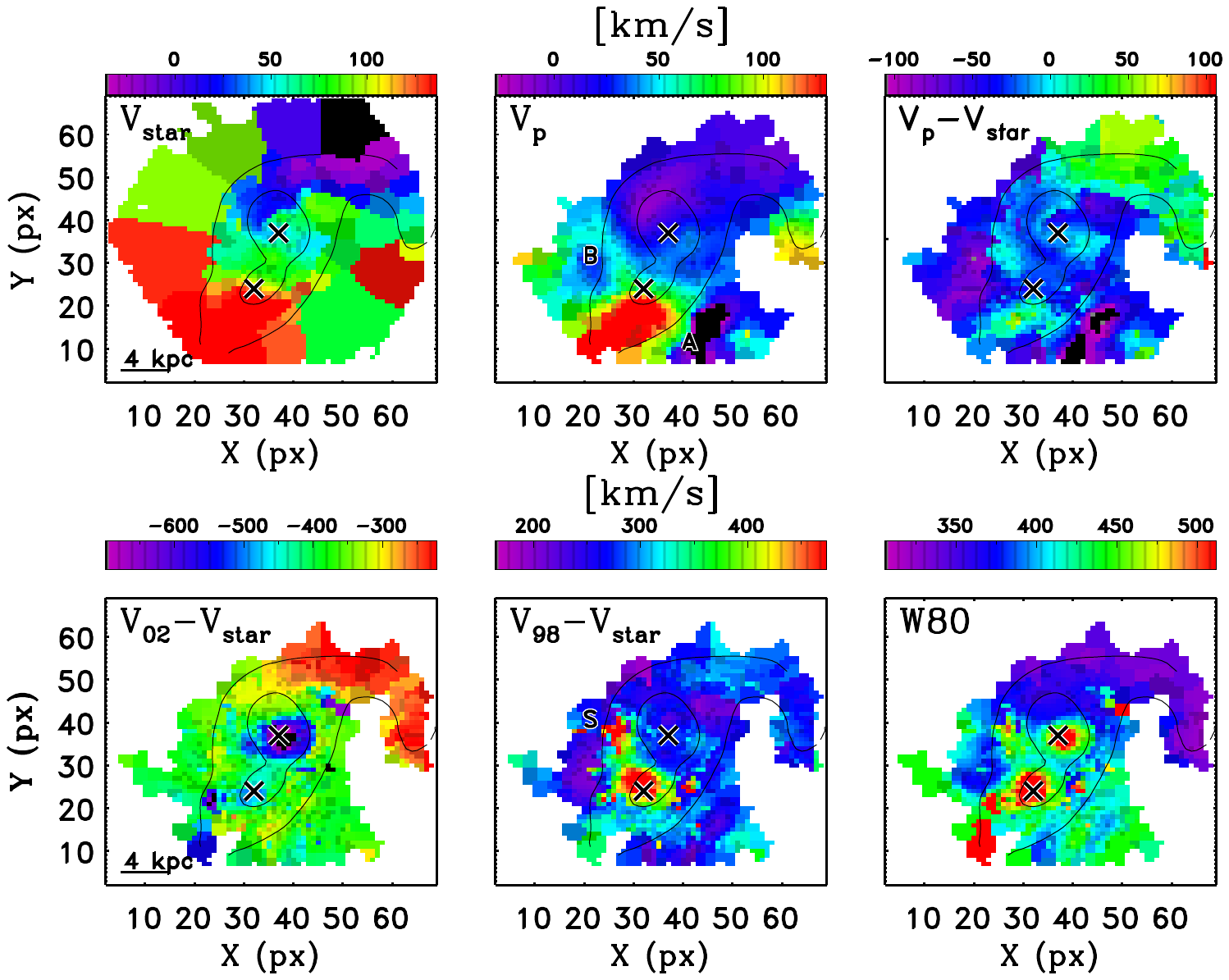}}

\caption{\small 
{\it Left panel:} Stellar velocity map derived from pPXF best-fit models. {\it Central panel:} [O III] non-parametric peak velocity $V_p$ estimate (an identical map is obtained for the H$\alpha$ line). {\it Right panel:} Difference between [O III] peak velocity ($V_p$) and stellar velocity (V$_{star}$); the most significant discrepancy between the two kinematic tracers is found in regions A and B in the central panel. All velocities are derived taking the narrow-component velocity of the $\mathcal{NW}$ -integrated spectrum as a reference. 
}

\label{velocitymap}
\end{figure*}

To derive spatially resolved  kinematic and physical properties of ionised and atomic gas, we analysed the spectra of individual regions in the FOV. Before proceeding with the fit, we smoothed the data-cube using a Gaussian kernel with an FWHM of three pixels ($1.5''$), taking into account the seeing conditions at the time of MaNGA observations (between $1.3''$ and 1.9$''$). A Voronoi tessellation was also derived to achieve a minimum S/N $= 10$ of the [O {\small III}]$\lambda$5007 line for each spaxel. We then applied BIC selection in each Voronoi bin to determine where a multiple-component fit was required to statistically improve the best-fit model. This choice allowed us to use the more degenerate multiple-component fits only where they were really needed. In order to mitigate the degeneracies in our fit as well as possible, we also considered an additional constraint for the relative fluxes of emission line features, requiring all the OC amplitudes to be smaller than those of NC (per individual line). This constraint was adopted because of the best-fit results obtained from the nuclear and outflow-integrated spectra; a visual inspection of individual fits and related residuals confirmed the advantages of using this.  

Figure \ref{Ngau} (left) shows a map reporting the number of sets of Gaussians we used to model the spectra in individual spaxels. As expected, most of the two- and three-set models were selected for the spaxels with the highest S/N, that is, in the regions close to the two nuclei\footnote{In the south-east region at the edge of the FOV, two Voronoi spaxels are associated with three-set models, because the intense H$\alpha$+[N {\small II}] system possibly shows asymmetric profiles. The [O {\small III}] lines are instead very faint in this region.}. The figure also displays the [O {\small III}]$\lambda$5007, H$\beta,$ and H$\alpha$ flux maps obtained by integrating the best-fit total emission line profiles. Some off-nuclear line emission is detected in the $\mathcal{NW}$ tidal tail and in the lower part of the figure, where we probably observe the overlapping of the two tidal tails that we identified in the HST images (T1 and T2 in Fig. \ref{HSTimages}, third panel).

Before introducing the velocity characterisation of the merging system, we show in Fig. \ref{OIIIspectra} a wavelength-collapsed image of the Mkn 848 high-$v$ [O {\small III}]$\lambda$5007 gas (absolute velocities $>250$ km/s in the $\mathcal{NW}$ nucleus reference frame). We also display a number of spectra extracted from $3\times 3$ spaxel regions. The spectra associated with the more nuclear regions display profiles that are more highly asymmetric, with prominent blue- and red-line wings, consistent with what we found in the integrated spectra (Sect. \ref{Sintegratedspectra}). The extracted spectra also highlight the variations in line peak wavelength across the FOV, and the possible presence of a double-peaked emission line at the overlapping of the T1 and T2 tidal tails (panel $f$). Broad profiles are found in the eastern quadrant above the $\mathcal{SE}$ nucleus, while narrower features are found in the spectra extracted from the T1 tail and the southern elongation. 

Figure \ref{NaIDspectra} shows the Na {\small ID} absorption and emission line fluxes across the FOV, obtained by integrating the spectral regions  $5860-5910\AA$ and $5890-5920\AA$, respectively, after ionised ISM and stellar contributions were subtracted.  
The maps show that the emission line contribution is small and is generally found at larger distances from the nuclear regions. We also display  different spectra extracted from distinct regions, in which Na {\small ID} is dominated by absorbing or emitting contributions. 

In the next sections we show that simultaneous fits successfully reproduce all the kinematic and flux variations described above for the ionised and atomic components. 
We anticipate, however, that the data quality and the complex broad profiles associated with the atomic and ionised gas tracers do not allow us to distinguish between the different outflow kinematic components that were used to reproduce the line profiles.

The pPXF analysis allowed us to trace the stellar velocities almost throughout the MaNGA FOV (Fig. \ref{velocitymap}, left). The stellar velocity map displays a strong gradient, from higher to lower velocities extending from $\mathcal{SE}$ to $\mathcal{NW}$ , taking as a reference the systemic of the $\mathcal{NW}$ nucleus; an inversion of the gradient is instead observed along the T1 tidal tail, which probably extends to the position of the $\mathcal{SE}$ nucleus (see the HST F160W image in Fig. \ref{HSTimages}). The extension of the T1 tail near the $\mathcal{SE}$ nucleus seems to have a stellar velocity that is similar to the velocities of the $\mathcal{SE}$ nucleus (see also Fig. \ref{OIIIspectra}). 

Similar velocity maps are obtained considering the $V_p$ of the [O {\small III}]$\lambda$5007 and H$\alpha$ lines (Fig. \ref{velocitymap}, central panel), suggesting that the velocity peak of ionised gas emission lines is a good tracer of the stellar motions, as a result of rotation in the individual galaxies and of the gravitational interaction (see e.g. the tidal tail T1). Because of the adopted constraints (Sect. \ref{Sintegratedspectra}), the velocity peak is always associated with the NC emission; therefore, we can reasonably assess that the OC does trace outflowing gas.   

Significant differences are found between $V_p$ and $V_{star}$  in the NE and SW regions around the $\mathcal{SE}$ nucleus, however (areas A and B in the central panel).
This discrepancy, highlighted in Fig. \ref{velocitymap} (right), is due to the significant variations in the $V_p$ velocities: in region A, the velocity shift can be explained by the interaction of tidal tails gas (see Fig. \ref{OIIIspectra}), while in region B, different kinematics may play a role (e.g. collimated outflow). In the following, we therefore use the stellar velocity map to subtract the gravitational motions and unveil the outflow kinematics.

The $V_p$  velocity maps in the vicinity of the nuclei are similar to those reported by \cite{Rupke2013}, who found a rotation pattern in the $\mathcal{SE}$ region and strong tidal motions and a near face-on orientation in the $\mathcal{NW}$ regions. 
In the next sections we discuss the kinematic properties of ionised and atomic gas by searching for deviations from the general motions traced by the stellar component and the bulk of ionised gas. In particular, we refer to the non-parametric maximum velocities ($V_{02}$ and $V_{98}$) of [O {\small III}] and Na {\small ID} features to trace the ejected gas kinematics.

\begin{figure*}[h]
\centering
\includegraphics[width=13.cm,trim=260 230 104 201,clip]{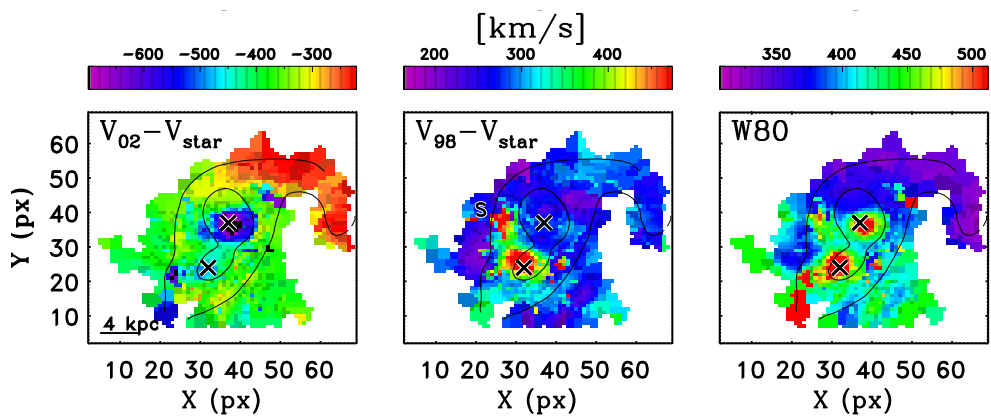}
\caption{\small 
Maps across the FOV of the maximum blue- (left) and red-shifted (central panel) velocities for the [O III] best-fit line profile. In particular, the two panels show the differences between maximum velocities ($V_{02}$ and $V_{98}$)  and stellar velocity (V$_{star}$), as labelled, to reveal outflow signatures. In the $V_{98}-V_{star}$ map the shell region (S) associated with both receding and approaching high-$v$ gas is shown. The right panel shows the [O III] velocity width $W80$.
}

\label{VmaxOIII}
\end{figure*}

\subsubsection{[O III] as a tracer of ionised outflow}\label{SOIIItracer}

Because of the strong velocity trend in the merging system, the maximum velocity maps mostly reproduce the same behaviour as the stars and the bulk of ionised gas (Fig. \ref{velocitymap}).
In order to unveil the outflows in the system, we therefore plot in Fig. \ref{VmaxOIII}  the difference between the non-parametric maximum velocities and $V_{star}$.  

In the $V_{02}-V_{star}$ map, velocities as high as $-550\div -650$ km/s are observed in the two nuclear regions, with projected extensions of $\sim 5$ kpc. The red-shifted maximum velocities ($V_{98}-V_{star}$)  instead display fast-receding gas in the proximity of the $\mathcal{SE}$ nucleus, with velocities of the order of $\sim 450\div 550$ km/s. 
Signatures of disturbed velocities are also found in the proximity of the region associated with the shell (S), with approaching as well as receding fast gas. In general, the distribution of high-$v$ gas suggests nuclear outflows in the two systems (see also Sect. \ref{multiphaseoutflow}).  

The H$\alpha$ velocity map confirms the presence and extension of fast gas in the proximity of the two nuclei.
We note, however, that although the H$\alpha$ line is stronger than that of doubly ionised oxygen, it might suffer from blending with [N {\small II}] emission lines. This means that the kinematics associated with Balmer emission might be affected by larger uncertainties. 

\subsubsection{Na {\small ID} as a tracer of neutral outflow}\label{SNaIDtracer}

\begin{figure*}[t]
\centering
\includegraphics[width=13.cm,trim=260 370 105 63,clip]{{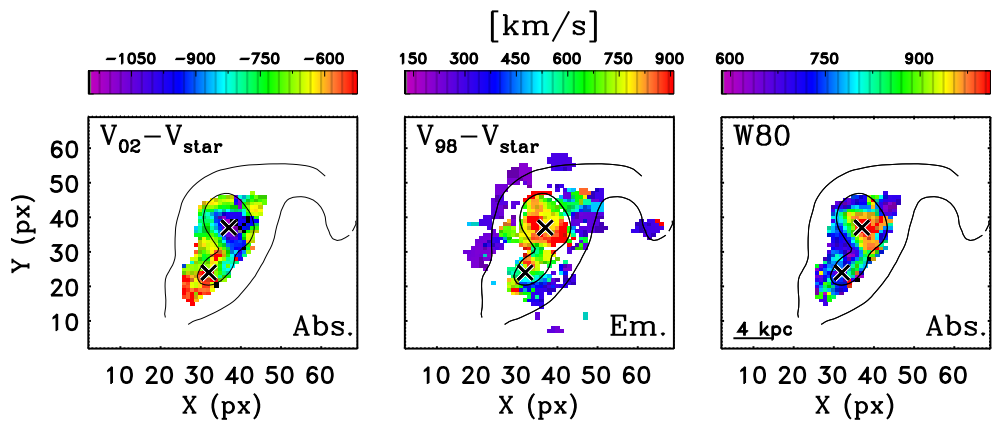}}

\caption{\small 
Sodium-line velocity maps: the maximum velocity $\Delta V_{02}$ (left panel), highlighting the receding-absorption components, and $\Delta V_{98}$ (central), highlighting the approaching-emitting components. The right panel shows the absorption component $W80$ velocity map. 
}

\label{VmaxNaID}
\end{figure*}

The Na {\small ID} negative maximum velocities were derived using the wavelength of the H transition of the $\mathcal{NW}$ integrated spectrum as a zero-point.  Figure \ref{VmaxNaID} shows the maximum velocity maps derived for the sodium absorption line. The negative $V_{02}-V_{star}$ tracer ($\Delta V_{02}$ hereinafter for simplicity) highlights strong neutral outflows in the $\mathcal{NW}$ nucleus (with $\Delta V_{02}\sim -1100$ km/s), and of slower winds with $\Delta V_{02} \sim -750$ km/s in the other nucleus.

Because of the geometrical configuration that is required to observe absorption features, we do not expect to observe the neutral counterpart of the red-shifted ionised outflows in absorption lines. On the other hand, red-shifted emission from the same transitions, as seen in the previous section, is detected throughout the FOV: this positive flux might be associated with the receding part of the outflow that is not visible in absorption, in the absence of significant absorption on the line of sight (LOS; \citealt{Prochaska2011}). 

The central panel in Fig. \ref{VmaxNaID} shows the $V_{98}-V_{star}$ ($\Delta V_{98}$ hereinafter) velocities derived for the sodium emission features. The highest velocities ($\sim 900$ km/s) are associated with the $\mathcal{NW}$  circum-nuclear regions; we note, however, that the low S/N (see Fig. \ref{NaIDspectra}) does not allow us to derive robust kinematics for this gas component. 
Moreover, as described in the previous section, modelling absorption and emission contributions simultaneously leads to significant degeneracy\footnote{Increasing the emission line flux can be compensated for in the $\chi^2$ minimisation by increasing the absorption line contribution.}. The simultaneous fit of all the emission and absorption lines allows us to mitigate this degeneracy: the sodium emission line of low-velocity gas is constrained to the same systemic as the ionised gas. 
We also note that the $\Delta V_{98}$ map qualitatively confirms that our fit models are able to mitigate the degeneracies in the Na {\small ID}, showing that the regions associated with very high-velocities are the same where ionised and neutral (absorbing) outflow material is also detected. 

MaNGA data allowed us to clearly detect high-$v$ sodium emission in Mkn 848 (e.g. Fig. \ref{NaIDspectra}). 
Resonant line emission from cool receding outflow gas has been observed in stacked spectra and in a handful of individual spectra (e.g. \citealt{Chen2010,Rubin2011}; see also e.g. \citealt{Martin2013} for Mg {\small II} resonant transition). To our knowledge, a spatially resolved resonant emission line from galactic wind has only been observed in two other systems: NGC 1808 (Phillips1993) and F05189-2524 (\citealt{Rupke2015}).
\subsubsection{Multi-phase outflow configuration}\label{multiphaseoutflow}

Figures \ref{VmaxOIII} and \ref{VmaxNaID} display disturbed kinematics in the proximity of  the  $\mathcal{NW}$  and  $\mathcal{SE}$ nuclei, as well as in the shell observed between the two systems. The disturbed kinematics involving a multi-phase medium is also highlighted by the [O {\small III}] and Na {\small ID} (absorption) $W80$ velocities that are reported in the same figures. 

Approaching outflowing ionised gas with velocities as high as $\sim -650$ km/s is observed in the $\mathcal{NW}$ regions. 
Neutral ejected material with maximum velocities of $\sim -1100$ km/s is also detected in the same regions. These values, in particular those related to the neutral component, are greater than the velocities generally found for SB-driven winds in local star-forming galaxies and ULIRGs (a few hundreds km/s; e.g. \citealt{Cazzoli2016,Chen2010,Rupke2013}), although this high-$v$ might be explained by smaller projection effects with respect to those of other SB-driven outflows. Instead, they are similar to the velocities found in AGN hosts (e.g. F06206-6315, \citealt{Cazzoli2016}; Mrk 273 and Mrk 231, \citealt{Rupke2013}). 
It is therefore possible that this material is ejected by AGN-driven winds, or at least SMBH activity might play a key role in this process. The presence of receding neutral high-$v$ gas is also detected in these nuclear regions (Fig. \ref{VmaxNaID}, central panel).

In the $\mathcal{SE}$ system we observe outflowing ionised emission associated with both approaching and receding material, with velocities of $\pm 500$ km/s. The spatial extent of this outflow and the near edge-on orientation of the  $\mathcal{SE}$ galaxy may suggest that the observed outflow fills a wide biconical region that probably encloses the high-$v$ gas close to the shell-like region. Approaching and receding neutral outflows, reaching velocities of $\Delta V_{max}\approx \pm 700$ km/s, are instead only revealed in the more nuclear regions because of observational limitations: the detection of absorbing Na {\small ID} is related to strong background emission.

Our spatially resolved spectroscopic analysis allows us to detect high-$v$ ionised gas out to a projected distance of $R_{out}^\mathcal{NW}([{\rm O\ {\small III}}]) \approx$ 4 kpc for the outflow associated with the $\mathcal{NW}$ nucleus, and of  $R_{out}^\mathcal{SE}([{\rm O\ {\small III}}]) \approx$ 6 kpc for the one related to the $\mathcal{SE}$ system (see Fig. \ref{VmaxOIII}). The observational limitations affecting the Na {\small ID} analysis allow us to derive lower limits on the real neutral gas widening, with $R_{out}^\mathcal{NW} ({\rm Na\ {\small ID}}) \gtrsim$ 4 kpc and $R_{out}^\mathcal{SE} ({\rm Na\ {\small ID}}) \gtrsim$ 2 kpc (Fig. \ref{VmaxNaID}).

In Sect. \ref{Soutflowproperties} we use these distances to derive outflow energetics assuming that the outflows extend out to these radii from the central regions.
We note that the near edge-on orientation of the $\mathcal{SE}$ galaxy suggests that the de-projected distances are not so different from the estimates reported before. On the other hand, the almost face-on configuration of the $\mathcal{NW}$ galaxy does not allow us to constrain the related outflow extension well, and large uncertainties might therefore be associated with these estimates. 

\begin{figure*}[h]
  \centering
  \includegraphics[width=13.cm,trim=122 250 240 180,clip]{{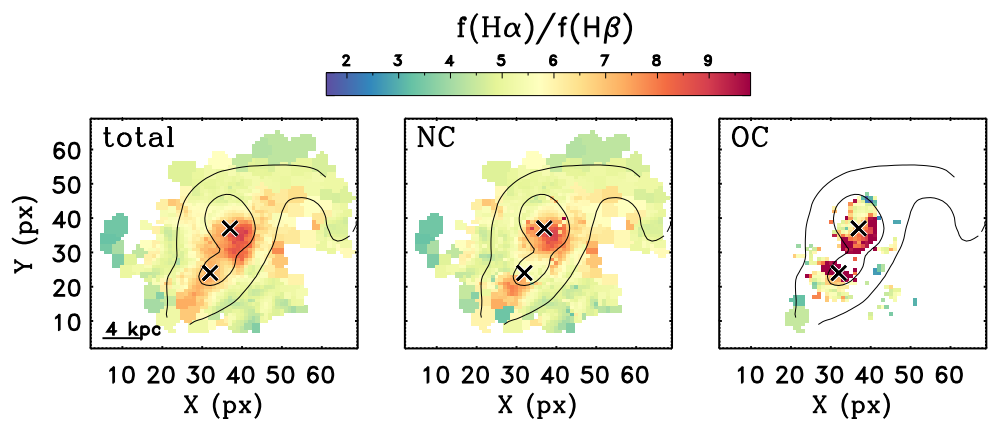}}
  \caption{\small 
Balmer decrement $f(H\alpha)/f(H\beta)$ ratio across the FOV for the total line profiles (left) and for the NC (centre) and OC (right) components separately. Higher flux ratios correspond to higher dust extinction.
   }
\label{extinction}
\end{figure*}

While the geometrical configuration of the $\mathcal{SE}$ galaxy and the detection of  multi-phase approaching and receding gas strongly suggest the presence of a biconical outflow whose axis is perpendicular to the galaxy major axis, the $\mathcal{NW}$ system could be more complex. 
Its almost face-on orientation would be compatible with a biconical outflow whose red-shifted gas component is mostly obscured by the host galaxy. Nevertheless, we traced faint high-$v$ Na {\small ID} gas emission, which is associated with receding gas. This cool material might suggest that the outflow in the $\mathcal{NW}$ galaxy has a wide aperture, such that the far parts of the cone that intercept the plane of the sky, are associated with positive velocities (e.g. \citealt{Perna2015a,Venturi2018}). Alternatively, this system might be associated with a more complex configuration between the outflow axis, the galaxy, and the plane of the sky (see e.g. \citealt{Fischer2013}). 
Data with a higher spatial resolution and S/N are required to distinguish between the different scenarios; for the purposes of this work, we assume that the two merging galaxies are associated with biconical outflows extending out to the distances $R_{out}$ reported above (but see \citealt{Rupke2013} for an alternative configuration of the  $\mathcal{NW}$ outflow). 

\section{Dust absorption}

Another key ingredient for characterising the outflow properties is the obscuration related to the ejected material. 
We used the Balmer decrement ratios $f(H\alpha)/f(H\beta)$ to estimate the reddening across the FOV.
A ratio of 3.1 (Case B, \citealt{Osterbrock2006}) is mostly used to determine the amount of extinction for low-density gas, such as that of the narrow-line regions (NLR). Any deviation to higher flux ratios is interpreted as due to dust extinction, which obscures emission lines at lower wavelengths. 

Figure \ref{extinction} shows the flux ratios separately for the total Balmer profiles and for NC and OC. The higher flux ratios, corresponding to the higher extinctions, are related to the nuclear positions and to the dust lane in the $\mathcal{SE}$ galaxy that is also visible in the three-colour \textit{HST} image (Fig. \ref{HSTimages}). The OC components are associated with even higher dust extinctions in the regions associated with the highest outflow velocities (Fig. \ref{VmaxOIII}). This is consistent with the measurements reported in the literature for AGN-driven outflows (e.g. \citealt{Holt2011,Mingozzi2018,Perna2015a,Villar2014}). However, such high dust extinctions might be the result of an incorrect  subtraction of the stellar component, which would mostly affect the H$\beta$ emission line wings (see e.g. \citealt{Perna2017a}). In Fig. \ref{outflowintegratedspectra} we show the MaNGA spectra and the best-fit pPXF spectra integrated over the outflow regions in the $\mathcal{NW}$ and $\mathcal{SE}$ galaxies. These integrated spectra show that the pPXF best fit does on average not underestimate the fluxes below the H$\beta$ wings (the best fit and the integrated spectra match almost perfectly, except for a negligible discrepancy in the red wings). We can therefore reasonably exclude that the high dust extinctions associated with the OC are due to an incorrect subtraction of the stellar component.

To explain these high extinctions, two possible explanations have been proposed: {\it i)} the outflows might be obscured by a dusty natal cocoon, or {\it ii)} ionised and dusty phases are mixed in the outflowing gas. Flux ratios above the 3.1 value are measured thorughout the entire FOV, which confirms the dust-obscured nature of the interacting system, but the highest flux ratios are spatially correlated with fast outflowing material. This argument, together with the fact that we observe both ionised and atomic outflows, points toward the scenario in which warm and dusty outflowing phases are mixed.

For the Case B ratio of 3.1 and the SMC dust-reddening law, the measured Balmer decrement ratios translate into a V-band extinction $A_V$ in the range between  1.3 and 3.1. These values are consistent with those derived from the integrated nuclear and outflow spectra presented in Sect. \ref{Sintegratedspectra} (Table \ref{spectralresults}), although the integrated measurements cannot account for the variations observed in Fig. \ref{extinction}. 

\section{Cool gas absorption and hydrogen column densities}\label{obscuration}

\begin{figure*}[t]
\centering
\includegraphics[width=18.cm]{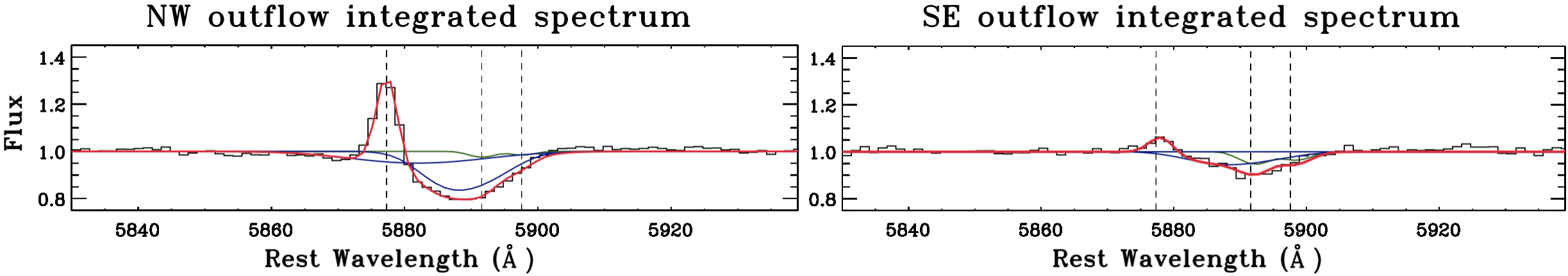}

\caption{\small 
Portion of the outflow-integrated spectra of Mkn 848 around the He I and Na ID lines for the $\mathcal{NW}$ ({\it left}) and $\mathcal{SE}$ ({\it right}) regions. The spectra are extracted from the spaxels associated with high-velocity gas in the neutral phase, with $\Delta V_{02} < -900$ km/s, for the $\mathcal{NW}$ system, and $\Delta V_{02} < -750$ km/s for the $\mathcal{SE}$ system. The two spectra have a normalised continuum, and the superimposed best-fitting  Na ID components in absorption are obtained following \citet{Rupke2005a}, assuming a geometrical configuration with partially overlapping sodium atoms (Sect. \ref{obscuration}). The He I emission line is instead fitted with a Gaussian profile. 
}

\label{rupkenaid}
\end{figure*}

In this section we derive the column density ($N_H$) of the wind, which is required to estimate how much neutral material is brought out in the outflow. The column density can be derived using the so-called doublet-ratio approach, which relates $N_H$  with the sodium equivalent width via the sodium curve of growth (e.g. \citealt{Cazzoli2016}). This approach assumes that the absorbing neutral material has a low optical depth ($\tau\lesssim 1$) and a uniform and total coverage of the continuum source, with a covering factor $C_f=1$. More plausible scenarios instead consider distributions of $\tau$ for individual kinematic components that also extend from low to high optical depths and $C_f \neq 1$  (e.g. \citealt{Rupke2005a} and references therein), as well as the possibility that sodium atoms at different velocities might be exposed to a different background continuum source (because of overlapping clouds on the LOS; see Fig. 6 in \citealt{Rupke2005a}). 

The very complex Na {\small ID} profiles and the low spectral resolution and quality of MaNGA data do not allow us to derive reliable spatially resolved column densities, especially because of the velocity overlap between the different kinematic components in the sodium profiles. Moreover, the multi-Gaussian fit performed in our analysis to derive the sodium kinematics instead assumes that all the kinematic components are co-spatial along the LOS and are exposed to the same background continuum source.

Because of these limitations, we decided to derive the average column densities associated with the neutral outflows in the $\mathcal{NW}$ and $\mathcal{SE}$ systems from the integrated spectra extracted from the spaxels associated with high-velocity gas in the neutral phase (see Sect. \ref{Sintegratedspectra}). These spectra, which are characterised by higher S/N, allowed us to follow a more rigorous approach. 
We fitted the observed intensity profiles of the sodium lines with a model parameterised in the optical depth space (e.g. \citealt{Rupke2002,Rupke2005a}), 

\begin{multline}\label{eqrupke}
I(\lambda) = 1- C_f \times [1-exp(-\tau_0 e^{-(\lambda - \lambda_K)^2/(\lambda_Kb/c)^2} - \\ 
2\tau_0 e^{-(\lambda - \lambda_H)^2/(\lambda_H b/c)^2})] ,
\end{multline}
where $I(\lambda)=f(\lambda)/f_0(\lambda)$ is the continuum-normalised spectrum, $C_f$ is the covering factor, $\tau_0$ is the optical depth at the line centre $\lambda_H$ ($5891\AA$), $b$ is the Doppler parameter ($b=FWHM/[2\sqrt{ln(2)}]$), and $c$ is the velocity of light. This model assumes that the velocity distribution of absorbing atoms is Maxwellian and that $C_f$ is independent of velocity. Following \citet{Rupke2005a}, we assumed the case of partially overlapping atoms on the LOS, so that the total sodium profile can be reproduced by multiple components and $I(\lambda)=\Pi_{i=1}^n I_i(\lambda) $, where $I_i(\lambda)$ is the $i$-th component (as given in Eq. \ref{eqrupke}) used to model the sodium features (see also Sect. 3.1 in \citealt{Rupke2002}). 

The best-fit models shown in Fig. \ref{rupkenaid} were obtained assuming one outflow component for the $\mathcal{SE}$ system and two outflow components for the $\mathcal{NW}$ galaxy, constraining their kinematics (FWHM and line centres) to the values previously obtained from the Gaussian fits (Fig. \ref{outflowintegratedspectra}). 
The existence of a multi-component outflow in the $\mathcal{NW}$ system is suggested by negative fluxes at $\approx 5870\AA$, next to the blue wing of the He {\small I} emission line. This contribution  is also clearly visible in Fig. \ref{NaIDspectra}, panel $b$, and in the SDSS legacy survey spectrum (\citealt{Perna2017a}). We associate  it with the Na {\small ID} absorption that also clearly affects the red wing of the He {\small I} line (see Figs. \ref{integratedspectra} and \ref{outflowintegratedspectra}).  The feature at $\approx 5870\AA$ suggests very high-$v$ gas ($V_{max} \approx -1200$ km/s), which is also confirmed  by \citet[][see their Table 3]{Rupke2013}. 
To obtain the best-fit models shown in Fig. \ref{rupkenaid}, we also constrained the covering factors and optical depths within the limits $0 \le \tau_0 \le 10^3$ and $0\le C_f \le 1$. 

We obtained for the outflow components covering factors of the order of $0.2-0.5$ and low optical depths ($\tau_0 \approx 0.4-0.8$). 
The column densities were computed from the best-fit values of $\tau_0$ and $b$ using Eq. 7 in \citet{Rupke2002}. 

Finally, following the prescriptions reported in \citet{Rupke2005b}, we derived the sodium abundance taking into account the ionisation fraction ($y=0.9$) and the depletion of Na atoms onto dust and the abundance as

\begin{equation}
log(N_{Na}/N_H)=log(1-y)+A+B,
\end{equation}
 where $A$ = log($N_{Na}/N_H$)$_{gal}$ is the Na abundance in the galaxy and $B =$ log($N_{Na}/N_H$) $-$ log($N_{Na}/N_H$)$_{gal} =-0.95$ is the depletion (the canonical Galactic value, \citealt{Savage1996}). Using the galaxy abundance derived from the Eq. 12 in \citealt{Rupke2005b}, 
we derived log($N_H/cm^{-2}$) $= 20\pm 0.1$ for  the $\mathcal{NW}$ and $\mathcal{SE}$ outflows\footnote{ The column density of the $\mathcal{NW}$ outflow was obtained by adding the $N_H$ of the two components that travel at different velocities. This result is not strongly affected by the assumption we used to model the sodium profile: forcing the fit to use one kinematic component, we obtained $C_f = 0.2 \pm 0.1$, $\tau_0 = 1\pm 0.2$ and a column density  log($N_H/cm^{-2}$) $= 19.8\pm 0.1$.}. These values are used to derive the atomic outflow energetics in Sect. \ref{Soutflowproperties}.

\section{Plasma properties}\label{Splasmaproperties}

Plasma properties represent further key ingredients to derive outflow energetics. 
Electron density ($N_e$) and temperature ($T_e$) within AGN-driven outflows can be derived using diagnostic line ratios involving forbidden line transitions (\citealt{Osterbrock2006}). At optical wavelengths, the [S {\small II}]$\lambda\lambda$6716,6731 and [O {\small III}] flux ratios (involving [O {\small III}]$\lambda\lambda$4959,5007 and [O {\small III}]$\lambda$4363)  have been used in the literature to derive the electron density and temperature for a limited number of sources (\citealt{Perna2017b} and references therein). These estimates are quite uncertain, however: doublet  ratio estimates for individual kinematic components are strongly affected by the degeneracies in the fit because the involved emission lines are faint and blended (see e.g. the detailed discussion in \citealt{Rose2018}). Moreover, the [S {\small II}] line ratio is sensitive to relatively low densities ($10^2\lesssim N_e/$cm$^{-3}$ $\lesssim 10^{3.5}$), while ionised outflowing gas has been found also in higher $N_e$ regions (e.g. \citealt{Lanzuisi2015,Villar2015}). Furthermore, the [O  {\small III}] and [S  {\small II}] species, being associated with different ionisation potentials, could arise from different emitting regions. 
However, while different line diagnostics have been proposed in the literature (e.g. involving trans-auroral emission lines; e.g. \citealt{Holt2011,Rose2018}), the [S {\small II}] line ratios still represent the more feasible diagnostic to derive the outflow electron density: the trans-auroral diagnostics  are very faint and can only be detected in local targets; moreover, the use of these diagnostics requires a wide spectroscopic wavelength coverage.

\begin{figure*}[t]
\centering
\includegraphics[width=13.cm,trim=105 250 230 180,clip]{{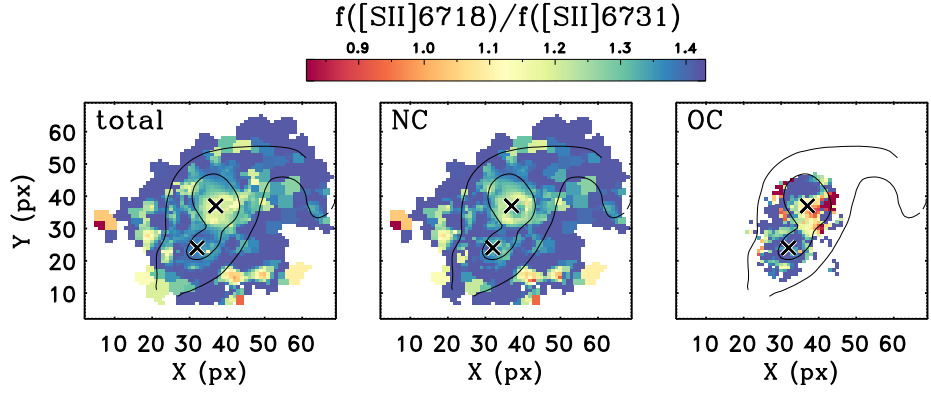}}
\caption{\small 
[S II] flux ratio maps derived for the total line profiles (left), and distinguishing between systemic (centre) and outflowing gas (right panel). Higher flux ratios correspond to lower electron densities.
}

\label{SIIratio}
\end{figure*}

In Fig. \ref{SIIratio} we show the [S {\small II}] ratio map, derived by computing the line ratio in each spaxel where the sulfur lines are detected with an S/N $>3$ for the total line profiles and for the NC and OC separately. The NC map shows that the lower ratios, which are associated with higher electron densities, are found in the proximity of the $\mathcal{NW}$ nucleus, with $N_e\sim 150\div 300$ cm$^{-3}$ (from Eq. 3 in \citealt{Perna2017b}, assuming $T_e=10^4$ K). The $\mathcal{SE}$ nuclear regions are instead associated with lower electron densities ($\lesssim 100$ cm$^{-3}$).

\begin{table*}
\footnotesize
\begin{minipage}[!hb]{1\linewidth}
\centering
\caption{Temperature- and density-sensitive line ratios from the $\mathcal{NW}$ and $\mathcal{SE}$ outflow-integrated spectra}
\begin{tabular}{lccc}
 Ratio  & &Value& Diagnostic \\
\toprule
\multicolumn{4}{c}{ $\mathcal{NW}$ outflow-integrated spectrum }\\
\hline
$[O {\small II}]\lambda$3727/$\lambda$3725 &  tot & $1.38\pm 0.01$ & $N_e = 122 \pm 12$ cm$^{-3}$ \\
 &  NC & $1.34 \pm 0.02$ &  $N_e = 180\pm 25$ cm$^{-3}$ \\ 
 &  OC & $1.45$  &  $N_e \approx 50$  cm$^{-3}$\\ 
$[S {\small II}]\lambda$6731/$\lambda$6716 & tot &  $1.16\pm 0.02$ & $N_e = 365\pm 21$  cm$^{-3}$\\ 
 & NC &  $1.26\pm 0.01$ & $N_e = 220\pm 20$  cm$^{-3}$\\ 
  & OC &  $0.89\pm 0.06$ & $N_e = 1080\pm 240$  cm$^{-3}$\\ 
$[$O {\small II}$](\lambda$3726+$\lambda$3729)/($\lambda$7319+$\lambda$7331) & tot &  $55 \pm 12$ & $N_e \lesssim 100$ cm$^{-3}$\\ 
$[$S {\small II}$](\lambda$4069+$\lambda$4077)/($\lambda$6716+$\lambda$6731) & tot &  $0.040 \pm 0.001$ & $''$\\ 
\cline{4-4}
$[O {\small III}](\lambda$5007+$\lambda$4959)/$\lambda$4363 &  tot & $64\pm 11$ & $T_e = 16000\pm 4000$\\ 
 & NC &  $83\pm 20$ & $T_e = 14350\pm 3800$\\ 
 & OC &  $49\pm 19$ & $T_e = 18600\pm 7060$\\ 
$[N {\small II}](\lambda$6548+$\lambda$6583)/$\lambda$5755 & tot &  $93\pm 25$ & $T_e = 10350\pm 1930$ K\\ 
 & NC &  $172\pm 7$ & $T_e = 8200\pm 75$\\ 
 & OC &  $53\pm 8$ & $T_e = 13600\pm 300$\\ 
$[$S {\small III}$](\lambda$9531+$\lambda$9069)/$\lambda$6312 & tot &  $325\pm 10$ & $T_e = 8970\pm 270$ K\\ 
\toprule
\multicolumn{4}{c}{ $\mathcal{SE}$ outflow-integrated spectrum }\\
\hline
$[$O {\small II}$]\lambda$3729/$\lambda$3726 &  tot & $1.37\pm 0.11$ & $N_e = 155 \pm 75$ cm$^{-3}$ \\
 &  NC & $1.45$ &  $N_e \approx 50$ cm$^{-3}$ \\ 
 &  OC & $1.35\pm 0.13$ & $N_e = 195\pm 90$ cm$^{-3}$ \\ 
$[S {\small II}]\lambda$6731/$\lambda$6716 & tot &  $1.36\pm 0.02$ & $N_e = 110\pm 25$ cm$^{-3}$\\ 
 & NC &  $1.32\pm 0.06$ & $150\pm 60$  cm$^{-3}$\\ 
  & OC &  $1.38\pm 0.04$ & $195\pm 45$  cm$^{-3}$\\ 
$[$O {\small II}$](\lambda$3726+$\lambda$3729)/($\lambda$7319+$\lambda$7331) & tot &  $74.5 \pm 2.7$ & $N_e \lesssim 100$ cm$^{-3}$\\ 
$[$S {\small II}$](\lambda$4069+$\lambda$4077)/($\lambda$6716+$\lambda$6731) & tot &  $0.108 \pm 0.005$ & $''$\\ 
\cline{4-4}
$[$O {\small III}$](\lambda$5007+$\lambda$4959)/$\lambda$4363 & tot &  $41.7\pm 0.1$ & $T_e =20400\pm 55$ K\\ 
$[$N {\small II}$](\lambda$6548+$\lambda$6583)/$\lambda$5755 & tot &  $160\pm 1$ & $T_e = 8430\pm 90$ K\\ 
$[$S {\small III}$](\lambda$9531+$\lambda$9069)/$\lambda$6312 & tot &  $>181^*$ & $T_e <6500$ K\\ 
\hline
\toprule
\end{tabular}
\label{plasmaproperties}
\end{minipage}
Note: De-reddened flux ratios (from Balmer decrement measurements) were used to the derive plasma properties reported in the table. 

$^*$: $3 \sigma $ upper limits. \\
\end{table*}

The outflow line ratios  are lower than those of NC, especially in the regions that are associated with the high-$v$ gas in the $\mathcal{NW}$ nucleus. They correspond to electron densities as high as $N_e \approx 1400$  cm$^{-3}$. The $\mathcal{SE}$ nucleus is instead associated with high line ratios that correspond to $N_e < 100$ cm$^{-3}$.

Because the [O {\small III}]$\lambda$4363 feature is so faint, we were unable to derive an electron temperature map from the [O {\small III}] ratio $f$($\lambda 5007 +\lambda 4959$)/$f$($\lambda$4363); we therefore used the outflow-integrated spectra obtained in Sect. \ref{Sintegratedspectra} to derive average electron temperatures associated with the outflowing gas. In Table  \ref{plasmaproperties} we report the $T_e$ and $N_e$ obtained from the integrated spectra; we note that the electron density values are consistent with the individual spaxel estimates (Fig. \ref{SIIratio}).

 The wide wavelength coverage of MaNGA observations allowed us to derive further estimates of the plasma properties from the $\mathcal{NW-}$ and $\mathcal{SE}$ -integrated spectra, in which other  faint temperature- and density-sensitive lines can be detected. In particular, we used the [S {\small II}] and [N {\small II}] temperature-sensitive line ratios, the [O {\small II}] density-sensitive ratio (\citealt{Osterbrock2006}), and the trans-auroral line diagnostic to derive $N_e$  (see Table \ref{plasmaproperties} for details). 
We modelled these additional emission lines using a multi-component simultaneous fit; 
for the features whose S/N was high enough to distinguish between different kinematic components, we derived temperature- and density-sensitive flux ratios for both unperturbed and outflowing ionised gas. 
All the line ratio diagnostics and the $T_e$ and $N_e$ measurements are reported in Table \ref{plasmaproperties}. 

We found that regardless of the small discrepancies between the values obtained with different tracers, systemic gas is in general associated with low electron densities (up to a few 100 cm$^{-3}$), while outflowing emission is associated with similar (in the $\mathcal{SE}$ regions) or higher densities (in the $\mathcal{NW}$ system; $N_e^\mathcal{NW}(OC) \approx 1500$ cm$^{-3}$). These results are consistent with previous works, in which both stacked spectra (\citealt{Perna2017b}) and high S/N spectra of individual targets (e.g. \citealt{Villar2015,Kakkad2018,Mingozzi2018,Rose2018}) have been analysed. 

Lower electron density values were instead derived from the trans-auroral line diagnostics, for which we were unable to distinguish between systemic and outflow components. The reason probably is that the systemic gas, associated with lower $N_e$, dominates the total line ratio fluxes. We also note that the oxygen-line doublet, being very close in wavelength, does not allow us to derive reliable measurements for $N_e$.

Electron temperatures of the order of $2\times 10^4$ K were derived from the [O {\small III}] line ratios for the systemic and outflowing gas, consistently with the results presented by \citet{Perna2017b}, for  the $\mathcal{NW}$ and $\mathcal{SE}$ systems (Table \ref{plasmaproperties}). Lower temperatures were instead measured from the sulfur and nitrogen diagnostics ($T_e\lesssim 10^4$ K). These discrepancies are typical of photoionised gas and can be explained considering the different excitation potentials of the different transitions: [N {\small II}] and [S {\small II}] are stronger in the outer parts of the photoionised regions, where the temperature is lower (\citealt{Osterbrock2006}; see \citealt{Curti2017}). In Sect. \ref{Soutflowproperties} we use the [S {\small II}]- and [O {\small III}]-based outflow plasma properties we derived from the $\mathcal{NW}$  and $\mathcal{SE}$ outflow-integrated spectra to constrain the energetics of the ejected material in the two systems. 

\section{Ionisation conditions}\label{ionisation}
\begin{figure*}[p]
\centering
\includegraphics[width=19.cm,trim=30 300 50 50,clip]{{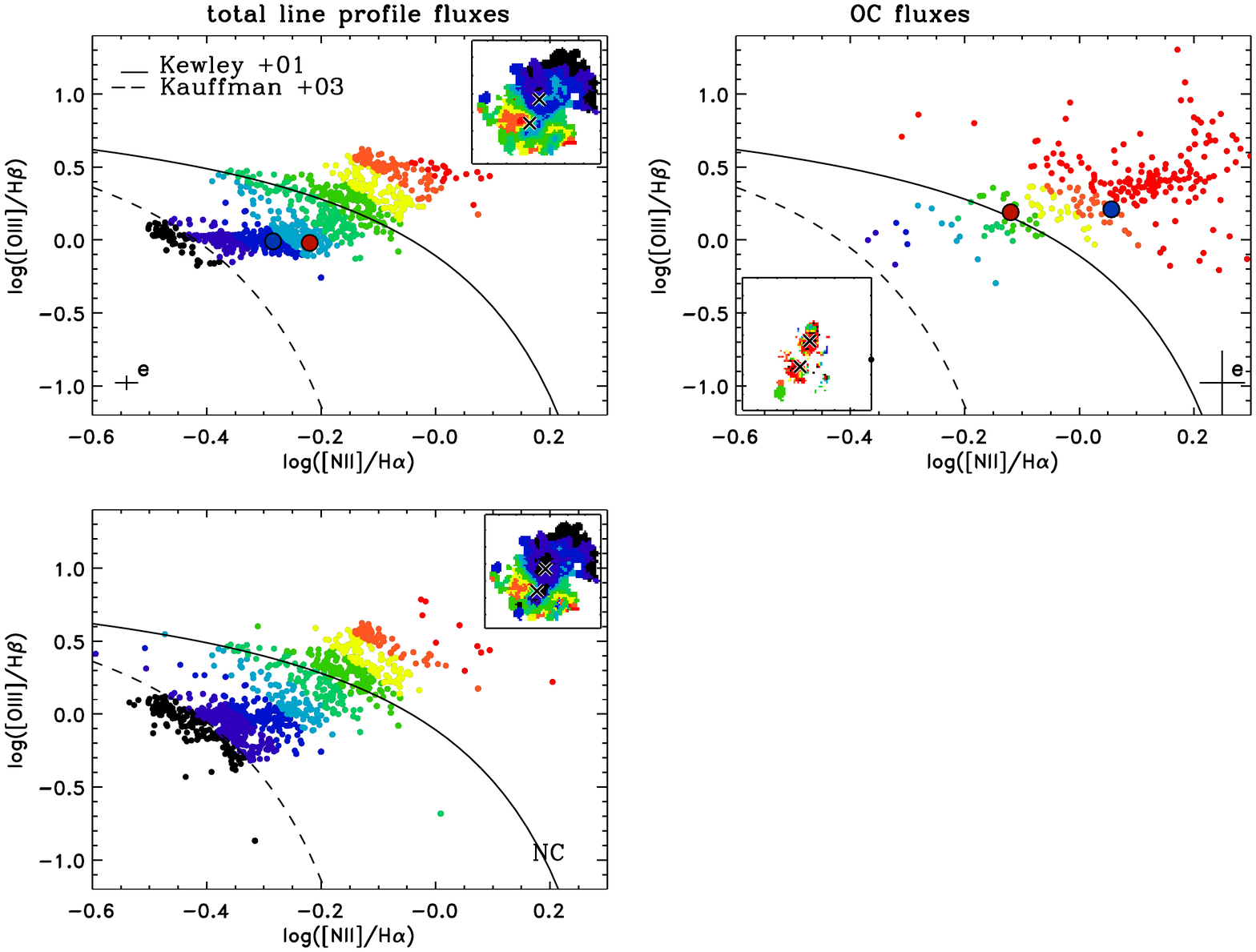}}
\includegraphics[width=19.cm,trim=30 300 50 70,clip]{{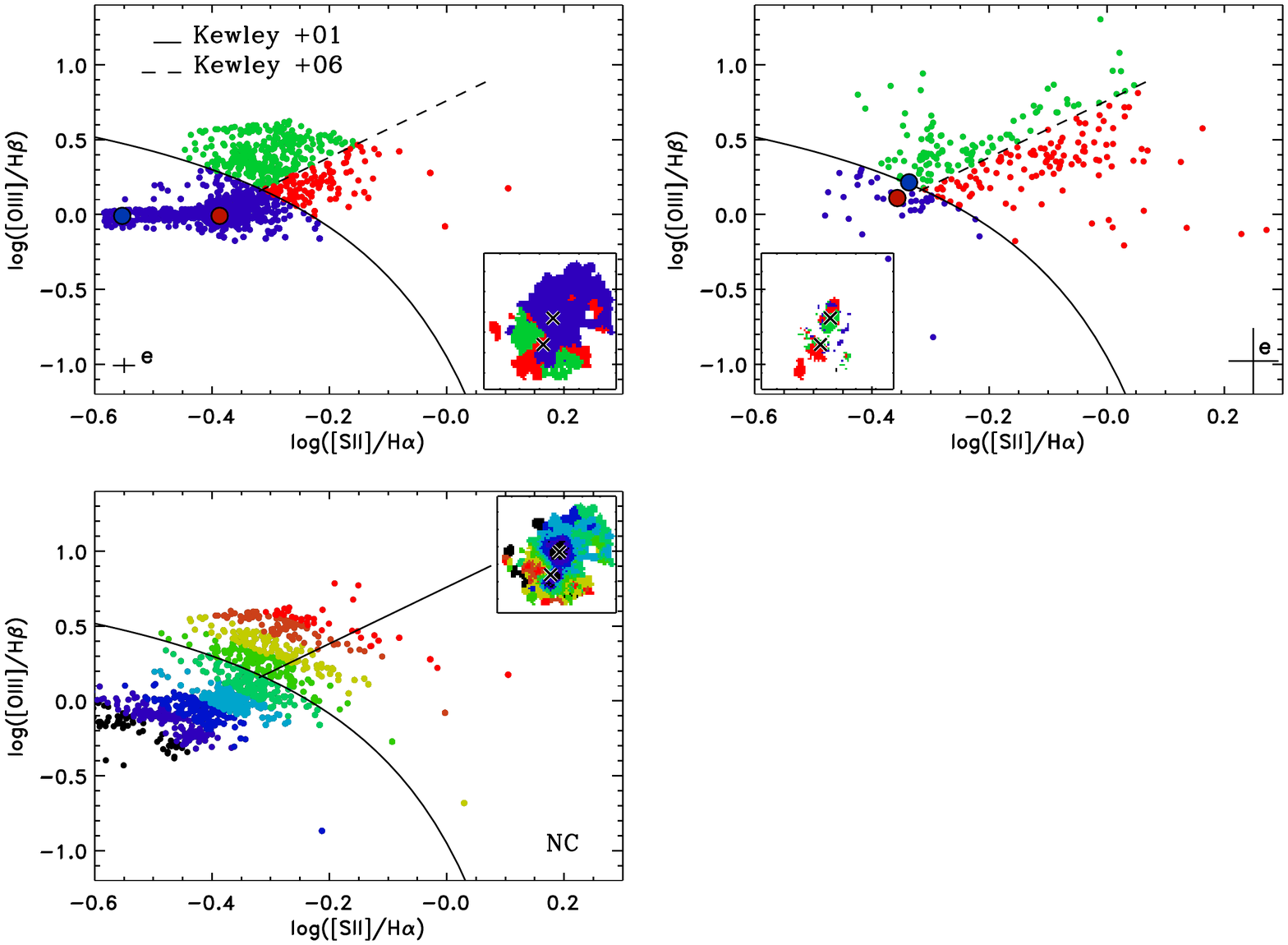}}
\includegraphics[width=19.cm,trim=30 300 50 70,clip]{{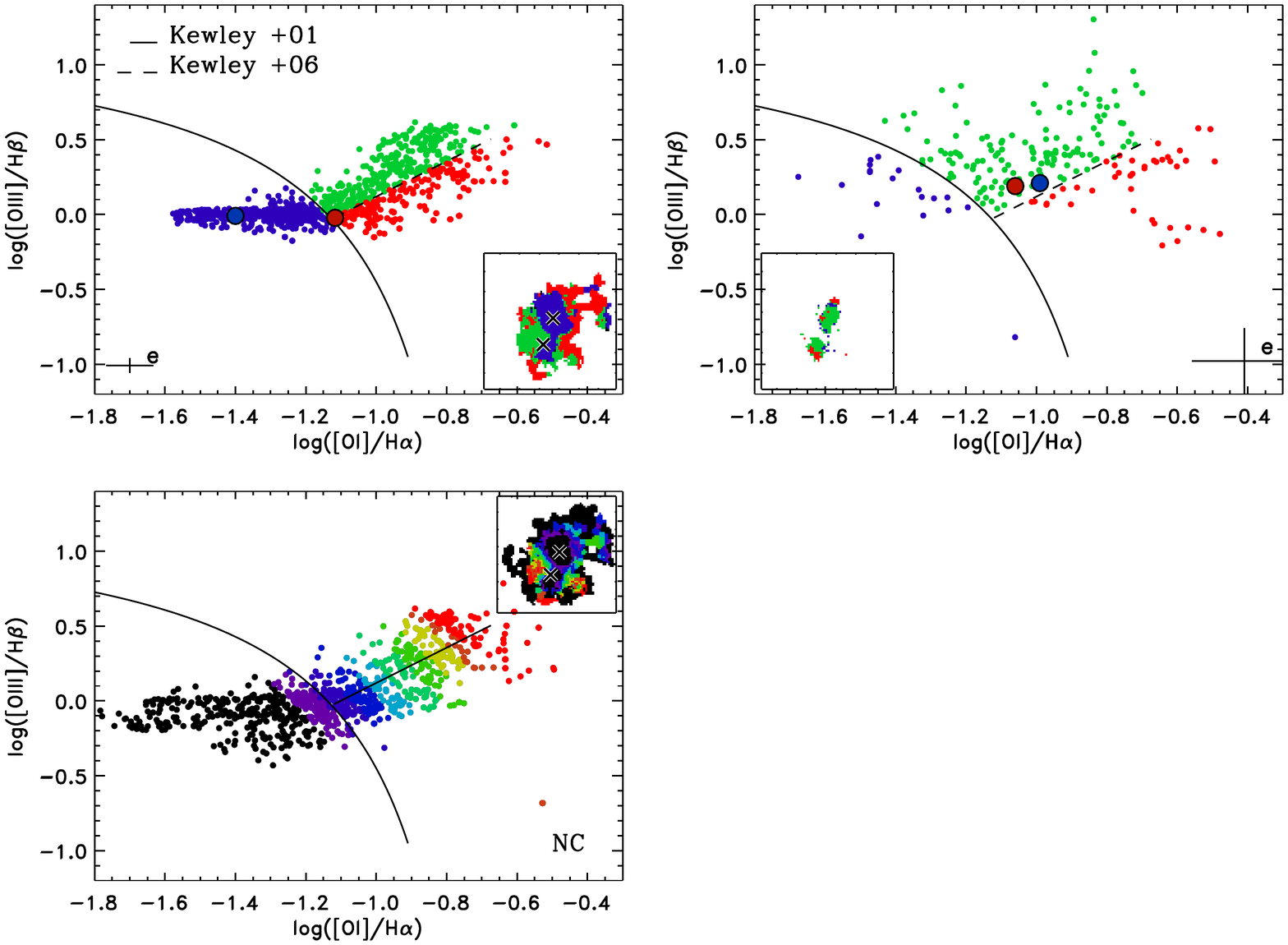}}
\caption{\small 
{\it Top panels:} Standard diagnostic [O III]/H$\beta$ vs. [N II]/H$\alpha$ diagrams. The lines drawn in the figures display the curves used to separate H II-like (below the dashed line) from AGN-ionised regions (above the solid line); the locus in between the two curves is associated with composite SF-AGN regions. The solid line is adapted from \citealt{Kewley2001}; the dashed curve is taken from \citet{Kauffman2003}. In the left panel, flux ratios are obtained considering the total line profiles (i.e. NC+OC fluxes); in the right panel, only the OC fluxes have been used. Small colour-coded dots refer to single spaxel measurements associated with the spatial regions shown in the insets (axis scales are the same as in all previous maps). Black spaxels are associated with flux ratios below the curve from \citet{Kauffman2003}; purple to red instead show the increasing distance from the same curve.
{\it Central panels:} Standard diagnostic [O III]/H$\beta$ vs. [S II]/H$\alpha$ diagrams. The lines separates  H II-like ionised regions (below the solid line) from AGN-ionised gas (above the dotted and dashed lines) and LINERs (in the right part of the diagrams). The solid and dashed lines are taken from \citet{Kewley2001} and \citet{Kewley2006}, respectively. Colour-coded flux ratio dots highlight the separation between AGN, LINERs, and H II regions, and refer to the spatial regions shown in the insets. 
{\it Bottom panels:} Standard diagnostic [O III]/H$\beta$ vs. [O I]/H$\alpha$ diagrams. The colours of the flux ratio dots and separation lines are the same as those in the central panels. 
In all figures, the largest blue and red dots refer to the flux ratios obtained from the $\mathcal{NW}$ and $\mathcal{SE}$ outflow-integrated spectra, respectively; only the flux ratios associated with H$\beta$ and [N II] (or [S II], in the central panels; or [O I] in the bottom panels) lines with an S/N $>3$ in are shown in the figures. Representative 1$\sigma$ errors are shown in the bottom corners of the plots.
}

\label{BPT}
\end{figure*}

In this section we investigate the dominant ionisation source for the emitting gas across the MaNGA FOV using the BPT (\citealt{Baldwin1981}) diagram. Figure \ref{BPT} (top panels) shows the [O {\small III}]$\lambda$5007/H$\beta$ versus [N {\small II}]$\lambda$6584/H$\alpha$ flux ratios, derived by integrating the line flux over the entire profiles (left panel) and the outflowing gas components (right), for only those spaxels in which all the emission lines (or the OC) are detected with an S/N $>3$\footnote{The representative errors in the figure are derived using Monte Carlo trials of mock spectra. Specifically, we randomly extracted 30 spectra from the data-cube and derived for each spectrum an estimate for the flux ratio errors, using 30 mock spectra (see e.g. \citealt{Perna2017a}). We then derived average errors from the individual errors associated with the 30 randomly extracted spectra.}. The curves drawn in the diagram correspond to the maximum SB curve (\citealt{Kewley2001}) and the empirical relation (\citealt{Kauffman2003}) used to separate purely SF galaxies from composite AGN-H {\small II} galaxies and AGN-dominated systems (e.g. \citealt{Kewley2006}). The points in the diagnostic diagrams are colour-coded based on the different subset selected in the BPT; the same colours are also reported in the map of Mkn 848 (insets in the figure) to distinguish the different ionisation sources in the FOV. 

The BPT diagram derived from total line profiles shows that the H {\small II}-like ionisation is found along the T1 tidal tail, and that the emission in the $\mathcal{NW}$ system is mostly associated with composite AGN-H {\small II} flux ratios. Both the strong star formation activity and the (possible) presence of a buried AGN (see Sect. \ref{ancillary}) in the nuclear region may therefore be responsible for the observed emission line ratios in the $\mathcal{NW}$ galaxy. As mentioned in Sect. \ref{multiphaseoutflow}, the observed outflow velocities can be used as a discriminant of SB- versus AGN-driven winds, pointing to the latter mechanism. However, it is possible that SB contributes to the outflow. Other diagnostics are therefore required to constrain the nature of the observed atomic and ionised outflows.

The ionisation conditions in the  $\mathcal{SE}$ system are instead more heterogenous: AGN-H {\small II} line ratios are observed only along the dust lane, while the directions perpendicular to the galaxy major axis are associated with AGN flux ratios. The presence of high-$v$ ionised and atomic gas in the same direction (see Figs. \ref{VmaxOIII} and \ref{VmaxNaID}) allows us to argue that the disturbed kinematics in the $\mathcal{SE}$ system are probably due to an AGN-driven outflow.

The BPT diagram derived for the OC emitting gas shows [N {\small II}]/H$\alpha$ ratios higher than those from NC+OC fluxes, with the highest values associated with the $\mathcal{SE}$ and $\mathcal{NW}$ nuclear regions. The shift to higher  [N {\small II}]/H$\alpha$ in outflowing gas is observed in the SF-driven (e.g. \citealt{Ho2014,Newman2012,Perna2018}) and AGN-driven (e.g. \citealt{McElroy2015,Perna2017b}) outflows. Therefore, it cannot be used to unambiguously distinguish between composite AGN-H {\small II} and AGN ionisation. Different mechanisms have been proposed to explain these higher flux ratios: they can be interpreted as due to shock excitation (e.g. \citealt{Allen2008}), hard ionisation radiation field (assuming that the outflowing material is close to the AGN; e.g. \citealt{Belfiore2016}), or they could be associated with high-metallicity regions within the outflow (\citealt{Villar2014}). The electron temperatures associated with outflowing gas (Table \ref{plasmaproperties}) are much lower than those expected for shock excitation ($T_e > 5\times 10^5$ K; \citealt{Osterbrock2006}), and may exclude the presence of shock excitation in Mkn 848. On the other hand, metallicity effects cannot be responsible for such high  [N {\small II}]/H$\alpha$ ratios (e.g. \citealt{Curti2017,Shapley2015}). 
We therefore suggest that a more plausible explanation for the higher [N {\small II}]/H$\alpha$ in the ejected gas could be given by hard ionisation radiation coming from the central SMBHs. 

The figure also shows the  [O {\small III}]$\lambda$5007/H$\beta$ versus [S {\small II}]$\lambda$6584/H$\alpha$ flux ratios (central panels) and [O {\small III}]$\lambda$5007/H$\beta$ versus [O {\small I}]$\lambda$6300/H$\alpha$ flux ratios (bottom panels). These diagrams also allow us to isolate the so-called low-ionisation nuclear emission line regions (LINERs; \citealt{Heckman1980}), which might be associated with shock excitation (e.g. \citealt{Allen2008}; but see \citealt{Belfiore2016}). The NC+OC flux ratios diagnostics confirm AGN-ionised gas in biconical regions in the $\mathcal{SE}$ system and H {\small II}-like regions in the $\mathcal{NW}$ source. The OC diagnostics indicate higher [S {\small II}]$\lambda$6584/H$\alpha$ and [O {\small I }]$\lambda$6300/H$\alpha$ ratios with respect to the NC+OC values; no clear spatial separation between AGN and shock excitation is found by comparing the different diagrams (see the insets in the panels), however. 

We also studied the correlation between the ejected gas kinematics, which are traced by the [O {\small III}] $W80$, and the ionisation state, which is traced by the [N {\small II}]/H$\alpha$, [S {\small II}]/H$\alpha$ and  [O {\small I}]/H$\alpha$  (\citealt{Dopita1995}; see e.g. \citealt{Arribas2014,Mingozzi2018,Perna2017a,Rich2015}). We found no clear correlation: all OC flux ratios seem to be independent of the gas kinematics, as expected for photo-ionised gas. Together with the arguments described above, this allows us to prefer AGN hard ionisation radiation over shocks in outflowing gas as a plausible explanation.

To summarise, the BPT diagnostics allow us to determine the nature of the ionised outflows in the AGN-dominated $\mathcal{SE}$ system; on the other hand, the $\mathcal{NW}$ nucleus has both a strong nuclear SB and an obscured AGN, and it is still not clear if the observed outflow can be unambiguously associated with AGN or SB activity, although the inferred velocities and the higher OC flux ratios seem to suggest that we observe AGN-driven outflows.

\section{Outflow energetics}\label{Soutflowproperties}

\begin{figure}[t]
\centering
\includegraphics[width=9.cm,trim=0 0 0 0,clip]{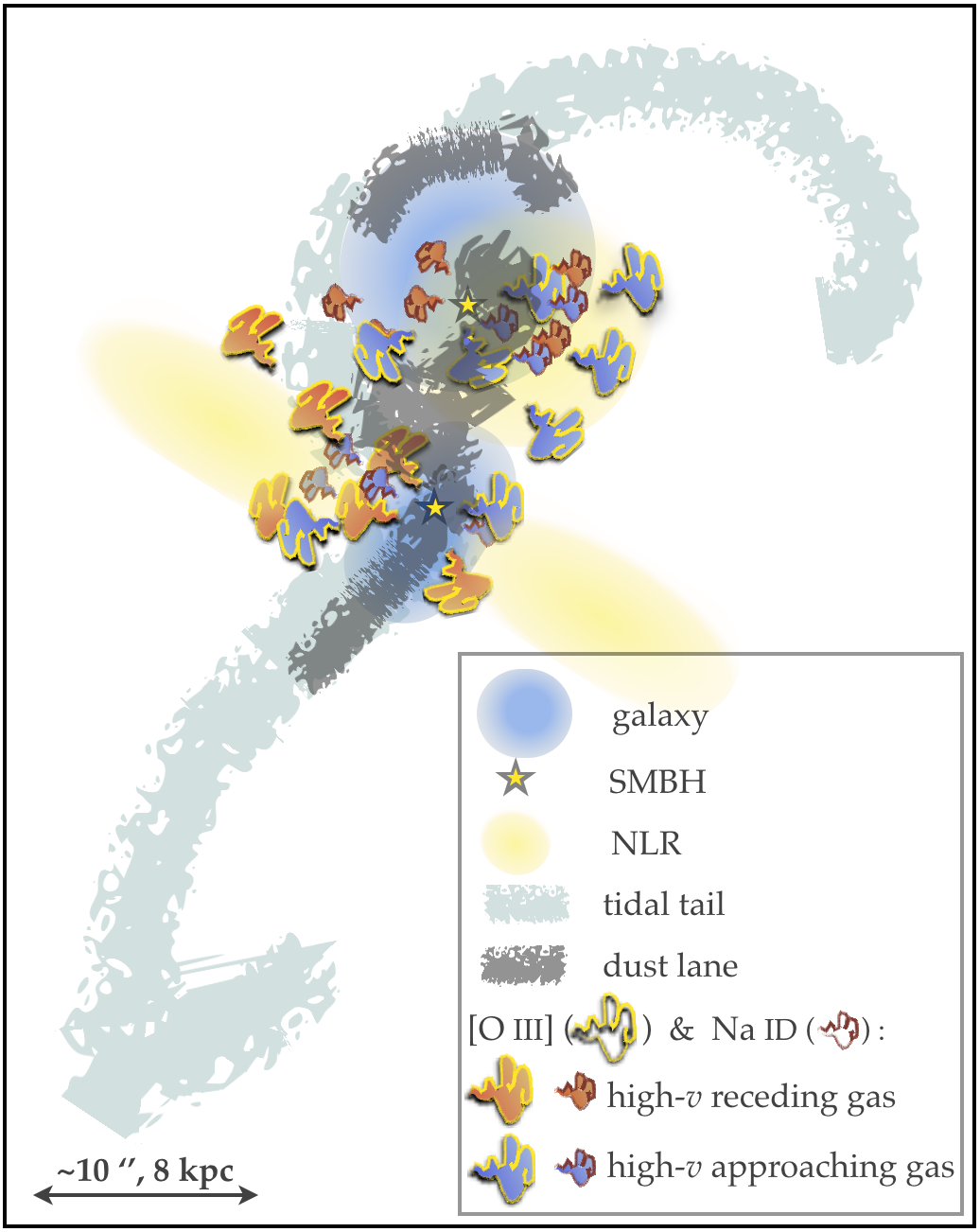}
\caption{\small 
Cartoon illustration for Mkn 848.  The blue ellipses and black and yellow stars represent the galaxies and the central SMBHs, respectively, with the near edge-on orientation for the $\mathcal{SE}$ system and the almost face-on configuration of the $\mathcal{NW}$ galaxy. The cyan and grey areas refer to the tidal arms and the dust observed in the \textit{HST} images.  
The extended and tenuous yellow area indicates the regions associated with AGN-ionisation: in the  $\mathcal{SE}$ system, we observe an extended NLR close to the plane of the sky; in the $\mathcal{NW}$, the almost-face configuration suggests that the NLR axis is close to the LOS.  The clouds indicate outflows and distinguish between ionised (yellow contour) and atomic (brown contour) gas, as well as between receding (red filled) and approaching (blue filled) material. Clouds visualise outflow directions, phases, and positions inferred from our analysis and do not refer to the (unknown) internal structure of the biconical outflows. The collected information suggest AGN-driven biconical outflows in the two merging galaxies.
}
\label{cartoon}
\end{figure}

Outflow mass rate, kinetic power, and mass-loading factor $\mu$, which is defined as the ratio between mass outflow rate and SFR, are important ingredients of galaxy evolution models (e.g. \citealt{OppenheimerDave2008,Hopkins2012}). These quantities have also been used to distinguish between AGN- and SF-driven outflow (e.g. \citealt{Brusa2015,Harrison2012}). In this section we report the ionised and atomic outflow properties associated with the ejected material observed in the two systems.
 
Taking advantage of the information so far collected regarding the spatial configuration of the two merging galaxies, the spatial distribution of high-$v$ atomic (Na {\small ID}) and ionised ([O {\small III}]) gas as well as the absorbing material (traced by Balmer decrement), and the ionisation conditions across the FOV, we show in Fig. \ref{cartoon} a cartoon illustrating the inferred properties of Mkn 848. For the ionised and atomic outflows, our findings are compatible with a simple biconical distribution (e.g. \citealt{Maiolino2012}). 

Following the arguments presented in \citet{Cresci2015}, we therefore computed the (biconical) outflow mass rates of the ionised components based on the equation

\begin{equation}\label{cresci}
\dot M^{out}_{H\beta}=8.6 \times \frac{L_{41}([H\beta])v_{out}}{N_{e} R_{kpc}} M_\odot/yr,
\end{equation}
where $L_{41}({\rm H\beta})$ is the H$\beta$ luminosity associated with the outflow component in units of 10$^{41}$ erg s$^{-1}$ and corrected for extinction, $N_{e}$ is the electron density, $v_{out}$ is the outflow velocity, and $R_{kpc}$ is the radius of the outflowing region in units of kiloparsec (kpc). The coefficient in Eq. \ref{cresci} was computed assuming $T_e=2\times 10^4$ K.

For comparison, we also computed the outflow mass rates of the ionised components from the equation presented by \citet{CanoDiaz2012}, who employed the [O {\small III}] luminosity:

\begin{equation}\label{canodiaz}
\dot M^{out}_{[{\rm O III}]}=0.48 \times \frac{CL_{41}([OIII])v_{out}}{N_{e} R_{out}  10^{\left [ O/H \right ]}} M_\odot/yr,
\end{equation}
where $L_{41}([OIII])$ is the [O {\small III}] luminosity associated with the outflow component in units of 10$^{41}$ erg s$^{-1}$ and corrected for extinction, $C$ is the condensation factor ($\approx$ 1), $N_e$ is the electron density, and 10$^{[O/H]}$ is the metallicity in solar units, $R_{out}$ is the radius of the outflowing region in units of kpc. The coefficient in Eq. \ref{canodiaz} was computed considering the dependence of the [O {\small III}] emissivity on the electron temperature ($2\times 10^4$ K)\footnote{The outflow mass rates we report in this work are $ \text{about three}$ times higher than those that would be obtained using the equation reported by \citet[][]{CanoDiaz2012} and derived assuming $T_e=10^4$ K.}.

The kinetic and momentum powers are derived from the relations 

\begin{equation}\label{EQkinetic}
\dot E^{kin}= \frac{1}{2} \dot M_{out} v_{out}^2 = 3.2\times 10^{35} \left( \frac{\dot M_{out}}{M_\odot /yr}\right ) \left( \frac{ v_{out}}{km/s} \right )^2  erg\ s^{-1},
\end{equation}

\begin{equation}\label{EQmomentum}
\dot P^{ion}= \dot M_{out} v_{out}  = 6.3\times 10^{30} \left( \frac{\dot M_{out}}{M_\odot /yr}\right ) \left( \frac{ v_{out}}{km/s} \right ) dyne.
\end{equation}

The outflow energetics were derived assuming a simple single radius wind because of the low spatial resolution and the unknown detailed geometrical configuration of the outflow.
All the quantities required to derive the outflow properties, as discussed in the previous sections,  are reported in Table \ref{outflowproperties} together with the outflow mass rates, and kinetic and momentum powers. The AGN ionisation does not allow us to derive metallicity measurements for the outflowing gas; therefore we  assumed a solar metallicity to derive [O {\small III}]-based outflow energetics. 
We note that these [O {\small III}]-based estimates  are $\approx 1$ dex lower than those obtained with H$\beta$, in agreement with previous results (see e.g. \citealt{Perna2015a,Carniani2015}). 
In the following, we therefore refer only to the H$\beta$-based outflow energetics for the ionised component of the multi-phase outflows.

The contribution to outflow energetics due to neutral ejected material was derived assuming the same geometrical configuration of ionised gas, starting from the equation (\citealt{Rupke2002})

\begin{equation}\label{MHout}
\dot M_{out}^H = 21 \left ( \frac{\Omega}{4\pi} \right ) C_f  \left ( \frac{N_H}{10^{21}cm^{-2}}\right ) \left (\frac{R_{out}}{1kpc} \right ) \left ( \frac{v_{out}}{200 km s^{-1}} \right )\ M_{\odot}\ yr^{-1}, 
\end{equation}
where $\Omega$ is the solid angle subtended by the wind, $C_f$ is the covering factor, $N_H$ is the column density, $R_{out}$ is the radius of the neutral outflow, and $v_{out}$ is the velocity of the ejected Na {\small ID}. The related kinetic and momentum powers were derived using Eqs. \ref{EQkinetic} and \ref{EQmomentum}. 

In deriving the atomic outflow energetics for the $\mathcal{NW}$ system, we assumed that the neutral outflow is related to the two distinct components travelling at different velocities\footnote{ This assumption in turn also involves those related to the adopted analytic functions that were used to model the Na {\small ID} system (Sect. \ref{obscuration}).} (Fig. \ref{rupkenaid}), and that $v_{out}$ is associated with the maximum velocity $V_{02}$ of a given component. 
All other measurements required to compute the outflow energetics and the derived values are reported in Table \ref{outflowproperties}. 

Our results suggest that the neutral and ionised outflows in the $\mathcal{SE}$ system are likely to have similar masses ($\dot M^{out}\approx 2\ M_\odot$/yr) and outflow energetics ($\dot E^{kin}\approx 10^{41}$ erg/s and $\dot P\approx 10^{33}$ dyne), while the $\mathcal{NW}$ outflow is dominated by the neutral component: the neutral mass outflow rate ($\dot M^{out}\approx 10\ M_\odot$/yr), kinetic ($\dot E^{kin}\approx 10^{42}$ erg/s), and momentum ($\dot P\approx 10^{35}$ dyne) powers are one or two dex higher than those associated with ionised gas. 
Discrepancies between ionised and atomic components have previously been reported for the $\mathcal{NW}$ system by \citet{Rupke2013}, and have also been found in other nearby systems (\citealt{Rupke2017}, $\sim 2\div 3$ dex; see also \citealt{Fiore2017}).
We also note that the outflow in the $\mathcal{NW}$ system is associated with higher mass rates and powers.

\citet{Rupke2013} detected both ionised and atomic outflows only in the $\mathcal{NW}$  system. With respect to their work, we used different techniques to derive outflow mass rates  and outflow energetics for the ionised and the atomic components. 
\citet{Rupke2013} used the H$\alpha$ emission to trace the ionised outflow, while we preferred to use the H$\beta$ emission, which is not affected by blending problems; they also assumed a very low electron density (10 cm$^{-3}$) to derive the ionised outflow properties, while we measured densities $> 1000$ cm$^{-3}$ in the $\mathcal{NW}$ outflowing gas. 
Our neutral outflow energetics are instead consistent with those derived in \citet{Rupke2013} within a factor of a few. 

In order to investigate the possible origin of the wind, we compared our inferred values of the total outflow kinetic power with the AGN bolometric luminosity and the expected kinetic power ascribed to stellar processes. The latter term was assumed to be proportional to the SFR, and is at most $\sim 7 \times 10^{41} \times$ (SFR [M$_\odot$/yr]) erg/s, following \citet{Veilleux2005}. We also derived the mass-loading factors associated with these outflows. SFR measurements were taken from \citet{Yuan2017}: SFR$^\mathcal{SE}=11$ M$_\odot$/yr and SFR$^\mathcal{NW}=76$ M$_\odot$/yr. The AGN bolometric luminosities of the two systems are only poorly constrained (Sect. \ref{ancillary}); we therefore decided to consider the (extinction-corrected, NC+OC) [O {\small III}] luminosity and a standard bolometric correction ($3\times 10^3$; \citealt{Heckman2004}) to obtain a rough estimate of the bolometric luminosities of the $\mathcal{NW}$ and $\mathcal{SE}$ AGN: log($L_{bol}(\mathcal{NW})$)$\approx 45.3$ erg/s and  log($L_{bol}(\mathcal{SE})$)$\approx 44.2$ erg/s\footnote{The  [O {\small III}]-based $\mathcal{NW}$ bolometric luminosity is $\sim 2$ dex higher than the luminosity obtained by X-ray analysis (Sect. \ref{ancillary}); we note, however, that the X-ray $L_{bol}$ should be considered as a lower limit because the $\mathcal{NW}$ nuclear region column density is poorly constrained and might be much higher than $N_H\approx 10^{24}$ cm$^2$.}. 

The outflow kinetic power of the $\mathcal{NW}$ and $\mathcal{SE}$ outflows can be associated with SF-driven winds, assuming a $\sim  10\%$ coupling between the stellar processes and the observed winds, and a mass-loading factor $\mu \approx 0.5$. The latter value is consistent with those derived for local luminous IR galaxies (\citealt{Arribas2014,Chisholm2017,Cresci2017}) and high-z star-forming galaxies (\citealt{Genzel2014,Newman2012,Perna2018}) without evidence of an accreting SMBH. 
 Similarly, the AGN luminosities in the two systems are high enough to sustain the inferred outflow powers (Table \ref{outflowproperties}): the value of $\dot E_{kin}$ is a few 0.1\% of the AGN bolometric luminosity for the $\mathcal{SE}$ and $\mathcal{NW}$ systems, in agreement with models, which predict a reasonable coupling between the energy released by the AGN and the energy required to drive the outflow (e.g. \citealt{King2005,Ishibashi2018}; see also \citealt{Harrison2018}).
These outflow energetics therefore do not allow us to unambiguously distinguish between SB- and AGN-driven outflows.

\begin{table}
\footnotesize
\begin{minipage}[!h]{1\linewidth}
\setlength{\tabcolsep}{1.8pt}
\centering
\caption{Outflow properties}
\begin{tabular}{lcc|c}
&$\mathcal{NW}$ & $\mathcal{SE}$ & Sect.\\
\toprule
\multicolumn{3}{c}{ionised component}&\\
\hline
$R_{out}$ (kpc)& 4 & 6 & \ref{multiphaseoutflow}\\
$N_e$ (cm$^{-3}$) & $1480\pm 870$ & $136\pm 45$&\ref{Splasmaproperties}\\
$T_e$ ($\times 10^4$ K) & $1.8\pm 0.4$ & $2.04\pm 0.01$&\ref{Splasmaproperties}\\
$v_{out}$ (km/s) & $-740\pm 20$ & $+460\pm 12$&\ref{SOIIItracer}\\
$L_{[OIII]}$ ($\times 10^{40}$ erg/s) & $5.9\pm 0.5$& $2.4\pm 0.3$&\ref{Soutflowproperties}\\
$\dot M^{out}_{[O\ III]}$ ($M_\odot$/yr) & $0.05\pm 0.03$&$0.07\pm 0.02$&''\\
$\dot E^{kin}_{[O\ III]}$ ($\times 10^{40}$ erg/s) & $0.8\pm 0.2$& $0.4\pm 0.1$&''\\
$\dot P_{[O\ III]}$ ($\times 10^{33}$ dyne) &$0.23\pm0.08$ & $0.20\pm 0.03$&''\\ 
$L_{H\beta}$ ($\times 10^{41}$ erg/s) & $1.4\pm 0.1$& $0.61\pm 0.09$&''\\
$\dot M^{out}_{H\beta}$ ($M_\odot$/yr) & $1.2\pm 0.7$&$3.0\pm 1.5$&''\\
$\dot E^{kin}_{H\beta}$ ($\times 10^{41}$ erg/s) & $1.8\pm 1.3$& $2.0\pm 1.3$&''\\
$\dot P_{H\beta}$ ($\times 10^{33}$ dyne) &$5.1\pm2.5$ & $4.4\pm 2.2$&''\\ 
\hline
\multicolumn{3}{c}{neutral component}&\\
\hline
$R_{out}$ (kpc)& $\gtrsim 4$ & $\gtrsim 2$& \ref{multiphaseoutflow}\\
$\Omega$  & $2\pi$* & $2\pi$* & \\
$v_{out}$ (km/s) & $(-550\pm 35);\ (-1150\pm 105)$ & $-530\pm 30$&\ref{SNaIDtracer}\\
$C_f$ & $(0.46 \pm 0.15);\ (0.15\pm 0.11)$ & $0.5\pm 0.2$ & \ref{obscuration}\\
log($N_H$) & $(20.0\pm    0.1) ;\ (19.9 \pm 0.1)$&$20.0 \pm 0.1$&\ref{obscuration}\\
$\dot M^{out}$ ($M_\odot$/yr) &$(5\pm 4);\ (17.5\pm 3)$&$5\pm 4$&\ref{Soutflowproperties}\\
$\dot E^{kin}$ ($\times 10^{41}$ erg/s) & $(4\pm 3);\ (70 \pm 25)$& $4\pm 2$&''\\
$\dot P$ ($\times 10^{33}$ dyne) &$(15\pm 13);\ (123\pm 15)$ & $15\pm 11$&''\\ 
\hline
\multicolumn{3}{c}{total (neutral+ionised)}&\\
\hline
$\dot M^{out}$ ($M_\odot$/yr) &$24\pm 5$&$8\pm 4$&\ref{Soutflowproperties}\\
$\dot E^{kin}$ ($\times 10^{41}$ erg/s) & $76\pm 25$& $6\pm 2$&''\\
$\dot P$ ($\times 10^{33}$ dyne) &$143\pm 20$ & $19\pm 10$&''\\ 
$\mu$ & 0.3& 0.7&''\\
$\dot E_{kin}/\dot E_{kin}^{SF}$ & 0.17& 0.08&''\\
$\dot E_{kin}/L_{bol}$ & 0.004& 0.004 &''\\
$\dot P/(L_{bol}/c)$ & $2$& $3$&''\\
\hline
\toprule
\end{tabular}
\label{outflowproperties}
\end{minipage}
Notes: [O III] and H$\beta$ luminosities are corrected for extinction using the values reported in Table \ref{spectralresults}. 
For the neutral phase of the  $\mathcal{NW}$ outflow, we report the derived properties for each of the two kinematic components described in Sect. \ref{obscuration}.
Mass-loading factors ($\mu$) and kinetic power expected for stellar processes ($\dot E_{kin}$)  are derived using the SFR reported in \citet{Yuan2017}. AGN bolometric luminosities from [O III] line flux: $L_{bol}^\mathcal{NW}\sim 10^{45.3}$ erg/s, $L_{bol}^\mathcal{SE}\sim 10^{44.2}$ erg/s. The asterisk indicates the assumed value for the solid angle of the wind.
Only statistical errors are reported for the outflow energetics (larger systematic uncertainties, e.g. due to an incorrect assumption on geometrical configurations, might affect our individual measurements; see e.g. \citealt{Perna2015b}).

\end{table}

\section{Conclusions}

We presented a spatially resolved spectroscopic analysis of Mkn 848, which is a complex system of two merging galaxies at z $\sim 0.04$. We analysed the publicly available MaNGA data to study the kinematic of warm and cool gas in detail. We distinguished between gravitational motions and outflow signatures and revealed for the first time multi-phase galaxy-wide outflows in the two merging galaxies. In particular, we revealed ionised and neutral ejected gas in the $\mathcal{NW}$ and $\mathcal{SE}$ systems; we also detected the Na {\small ID} emission across the MaNGA FOV, which is also associated with receding outflow gas in the two merging galaxies.

We tested whether the ejected material can be associated with AGN winds or stellar processes, taking advantage of a detailed analysis of the ionised and atomic gas kinematic, ionisation conditions, and outflow energetics. 
The energetics of the $\mathcal{NW}$ and $\mathcal{SE}$ outflows did not allow us to unambiguously distinguish between SB- and AGN-driven outflows: the derived mass-loading factors and outflow kinetic powers would require a $\sim 10\%$ coupling between stellar processes and the winds to explain the outflow measurements; much smaller coupling factors would instead be required in case of AGN winds. Similarly, the BPT diagnostics did not allow us to definitely distinguish between AGN and SB ionisation. 
Nevertheless, the collected information might suggest an AGN origin of the observed multi-phase outflows: among them are the high-outflow velocities, of the order of 500-1000 km/s; the BPT flux ratios associated with the outflow components, significantly higher than those of systemic gas; and the absence of any evidence of shock ionisation.
Further deeper multi-wavelength observations are required to better constrain the nature of these multi-phase outflows, however.

Our analysis showed that detailed multi-wavelength studies are required to comprehensively characterise the outflows. We found that the neutral outflow component might be related to mass outflow rates and kinetic and momentum powers similar to or even higher than those of the ionised component. This result further emphasises  the need for follow-up observations aimed at detecting and characterising the different outflow phases in individual targets.

We finally note that evidence of vigorous and efficient outflows in early- or intermediate-stage merger systems has been collected for several nearby ULIRG/AGN galaxies (e.g. \citealt{Feruglio2013,Feruglio2015,Rupke2013,Rupke2015,Saito2017}) and high-z SMG/QSOs (e.g. \citealt{Perna2018b} and refs therein). These findings suggest that feedback phenomena could be important starting at the early phases of the SB-QSO evolutionary sequence, efficiently depleting the gas reservoirs in the systems (as observed in z $>1$ CT QSOs with gas fractions much lower than in normal galaxies with similar properties). This means that these results might be at odds with the scenario in which the important feedback phase follows the SF-dominated phases that are related to the initial merger stages (e.g. \citealt{Hopkins2012b}). 

\vspace{2cm}

{\small 
{\it Acknowledgments:} We acknowledge the anonymous referee for their constructive comments, which significantly contributed to improving the quality of the paper.
GC and AM acknowledges support by INAF/Frontiera through the `Progetti Premiali' funding
scheme of the Italian Ministry of Education, University, and Research. GC, AM and FM acknowledges support by INAF PRIN-SKA 2017 programme 1.05.01.88.04. 
 We acknowledge financial support from INAF under contract PRIN-INAF-2014
(``Windy Black Holes combing Galaxy evolution''). 
Part of this research was carried out at the Munich Institute for Astro- and Particle Physics (MIAPP) of the DFG cluster of excellence ``Origin and Structure of the Universe'' during the program ``In \& Out: What rules the galaxy baryon cycle?'' (July 2017). 
MP thanks M. Dambrosio, who provided a useful and rare insight into 3D projection.

Funding for the Sloan Digital Sky Survey IV has been provided by the Alfred P. Sloan Foundation, the U.S. Department of Energy Office of Science, and the Participating Institutions. SDSS-IV acknowledges
support and resources from the Center for High-Performance Computing at
the University of Utah. The SDSS web site is www.sdss.org.

SDSS-IV is managed by the Astrophysical Research Consortium for the 
Participating Institutions of the SDSS Collaboration including the 
Brazilian Participation Group, the Carnegie Institution for Science, 
Carnegie Mellon University, the Chilean Participation Group, the French Participation Group, Harvard-Smithsonian Center for Astrophysics, 
Instituto de Astrof\'isica de Canarias, The Johns Hopkins University, 
Kavli Institute for the Physics and Mathematics of the Universe (IPMU) / 
University of Tokyo, Lawrence Berkeley National Laboratory, 
Leibniz Institut f\"ur Astrophysik Potsdam (AIP),  
Max-Planck-Institut f\"ur Astronomie (MPIA Heidelberg), 
Max-Planck-Institut f\"ur Astrophysik (MPA Garching), 
Max-Planck-Institut f\"ur Extraterrestrische Physik (MPE), 
National Astronomical Observatories of China, New Mexico State University, 
New York University, University of Notre Dame, 
Observat\'ario Nacional / MCTI, The Ohio State University, 
Pennsylvania State University, Shanghai Astronomical Observatory, 
United Kingdom Participation Group,
Universidad Nacional Aut\'onoma de M\'exico, University of Arizona, 
University of Colorado Boulder, University of Oxford, University of Portsmouth, 
University of Utah, University of Virginia, University of Washington, University of Wisconsin, 
Vanderbilt University, and Yale University.

}

\end{document}